\documentclass[lettersize,journal]{IEEEtran}
\usepackage{amsmath,amsfonts}
\usepackage{amsmath,amssymb,amsthm}

\usepackage{algorithmic}
\usepackage{algorithm}
\usepackage{array}
\usepackage[caption=false,font=normalsize,labelfont=sf,textfont=sf]{subfig}
\usepackage{textcomp}
\usepackage{stfloats}
\usepackage{url}
\usepackage{verbatim}
\usepackage{booktabs}
\usepackage{siunitx}
\usepackage{amssymb}
\usepackage{graphicx}
\usepackage{multirow}
\def\BibTeX{{\rm B\kern-.05em{\sc i\kern-.025em b}\kern-.08em
    T\kern-.1667em\lower.7ex\hbox{E}\kern-.125emX}}
\usepackage{balance}
\begin{document}
\title{Critical-Set-Aided Simplified Blind SCL Recognition of Polar Codes}
\author{Changwei Tu, Cheng Yang, Xianzhao Feng, and Kai Niu
\thanks{This work is supported by the National Natural Science Foundation of China under Grant 62321001 and Grant 62471054.
  \emph{(Corresponding author: Kai Niu.)}
  
The authors are with the Key Laboratory of Universal Wireless Communications, Ministry of Education,
Beijing University of Posts and Telecommunications, Beijing, China.
Email: \{tuchangwei, yscheng, foreseexz, niukai\}@bupt.edu.cn.}
  }

\markboth{Journal of \LaTeX\ Class Files,~Vol.~18, No.~9, September~2020}%
{How to Use the IEEEtran \LaTeX \ Templates}

\maketitle

\begin{abstract}
Blind recognition of polar codes from noisy observations is a key problem in non-cooperative signal processing.
Although existing blind successive cancellation list (BSCL) recognition exploits channel soft information, it performs two-hypothesis path expansion at every source-bit position, resulting in high complexity. 
In this paper, we first analyze the first recognition-error positions in the blind successive cancellation (BSC) recognition 
and observe that they are closely related to the corresponding contribution terms in the existing Bhattacharyya-parameter-based upper bounds.
Based on this observation, a critical-set-aided simplified blind successive cancellation list (SBSCL) recognition method is proposed. SBSCL performs two-hypothesis path expansion only at the selected critical-set positions and keeps BSC recognition at the remaining positions, thereby reducing complexity.
To improve the reliability of critical-set selection and refine the performance analysis, density-evolution (DE)-based bounds are further developed. 
Under the ideal SC-consistent condition,
the synthetic log-likelihood-ratio (LLR) distributions obtained from density evolution are used to compute the optimized Chernoff coefficient for the upper bound and the overlap coefficient for the lower bound.
Simulation results show that the DE-based bounds are tighter than the Bhattacharyya-parameter-based bounds. In the considered settings, the gap between the DE upper and lower bounds is within $1$ dB around a recognition-error probability of
$10^{-2}$. Furthermore, SBSCL achieves nearly the same recognition success rate as BSCL, and the size of critical set decreases rapidly as the signal-to-noise ratio (SNR) increases.

\end{abstract}


\section{Introduction}

\IEEEPARstart{I}{n} non-cooperative signal processing, intercepted noisy observations are often analyzed with limited prior information about the transmitter [1]. Key transmission parameters, such as the modulation format, frame structure, and coding information, may therefore be unknown and must be inferred from the received data [2], [3]. For coded transmissions, the hidden coding structure includes the coding type, code length, code rate, and construction parameters. Identifying such coding information from  received data
is referred to as blind channel-code identification in [4].

Existing work on blind identification has considered several representative channel codes. 
For linear block codes with explicit algebraic structures, such as Bose–Chaudhuri–Hocquenghem (BCH) and Reed–Solomon (RS) codes, 
most methods rely on parity-check relations or the structural properties of generator polynomials [5]--[7].
For convolutional and Turbo codes, state-transition relations, trellis representations, and iterative decoding characteristics are often used to analyze the coding parameters [8], [9].
For Low-Density Parity-Check (LDPC) codes, recognition mainly relies on the sparse parity-check matrix and its low-density check relations [10], [11]. 
These methods are effective for their target codes, but they are not always easy to extend to other coding schemes because of the differences in code structures.

Polar codes, introduced by Ar{\i}kan, are constructed by exploiting channel polarization [12].
Due to their excellent decoding performance and low-complexity encoding and decoding algorithms [13], polar codes were selected as the coding scheme for 5G control channels [14]. 
The application of polar codes also raises the problem of blind recognition  under non-cooperative conditions [15].
In this case, blind recognition of polar codes mainly includes the estimation of the code length, code rate, and frozen-set or information-set pattern.

The blind recognition of polar codes has been investigated in several studies.
In [16], an information-matrix extension method was proposed to recognize the code length and the positions of information bits. 
Its decision threshold is determined by the channel error condition and the structure of the polar-code generator matrix. 
Based on the recursive inheritance properties of polar codes,  the authors of [17] introduced an estimation-and-derivation strategy, where the recognition result for shorter codes is used to assist the recognition of longer codes, thereby improving reliability. 
In [18], the recognition performance under high-BER conditions was further improved by introducing multi-threshold voting and partial-order supplementation.

The above methods mainly use hard-decision codewords and structural properties of polar codes, without directly using the soft information in the intercepted observations. 
In contrast, the authors of [19] and [20] introduced soft information to construct decision statistics and derive the corresponding decision thresholds.
However, their recognition performances are still limited. 
After that, the authors of [21] proposed a blind successive cancellation list (BSCL) method for polar-code recognition with known code length. 
In BSCL, each surviving path is expanded under the frozen-bit and information-bit hypotheses at each source-bit position, and the path is updated according to multiple intercepted observations. 
In [22], the decision metric and reference threshold of BSCL were interpreted from a hypothesis-testing viewpoint. 
Under the ideal successive-cancellation (SC) consistent condition, upper and lower bounds on the recognition error were further derived in terms of the Bhattacharyya parameters.

However, the Bhattacharyya-parameter-based bounds in [22] can be loose. 
In addition, BSCL has relatively high decoding complexity, since each surviving path is expanded under both the frozen-bit and information-bit hypotheses at every source-bit position. 
To address these issues, this paper develops a critical-set-aided simplified BSCL (SBSCL) recognition method and derives density-evolution (DE)-based bounds for performance analysis and critical-set selection. 
The key idea is to retain the two-hypothesis path expansion only at the source-bit positions that are more likely to cause recognition errors, while processing the remaining positions by a single local decision. 
The main contributions of this paper are summarized as follows:
\begin{itemize}
   \item We propose a critical-set-aided simplified BSCL method, referred to as SBSCL. 
   By analyzing the first recognition-error positions, we show that early errors are mainly concentrated at a small number of unreliable synthetic channels, whose indices constitute the critical set. 
   Accordingly, SBSCL selects a critical set offline based on the position-wise upper-bound contributions and performs the BSCL two-hypothesis path expansion only at the positions in this set.

  \item We introduce a DE-based recognition-error analysis framework under the ideal SC-consistent condition. Starting from the channel log-likelihood-ratio
  (LLR) distribution, density evolution is performed along the polarization transform to obtain the synthetic LLR distribution of each bit position. The resulting DE distributions are then used to characterize the frozen-bit and information-bit hypotheses, thereby providing a statistical interpretation of the recognition metric and the neutral decision threshold.

  \item Based on the DE distributions and the ideal SC-consistent condition, we derive upper and lower bounds on the recognition error. The upper bound is obtained from an optimized Chernoff coefficient, while the lower bound is obtained by the corresponding overlap coefficient. Compared with the Bhattacharyya-parameter-based bounds, the proposed bounds give a tighter evaluation and provide a more precise position-wise contributions used for critical-set selection.
   
\end{itemize}

The remainder of this paper is organized as follows. 
Section II presents the system model, SC recursion and synthetic LLRs, the BSCL recognition method, and density-evolution preliminaries. 
Section III analyzes the first recognition-error positions and introduces the  SBSCL recognition method. 
Section IV derives the DE-based  bounds. 
Section V presents simulation results for the theoretical bounds, recognition performance, and complexity reduction. 
Section VI concludes the paper.

\section{Preliminaries}

\subsection{System Model}

Let $\mathcal{C}(N,K)$ denote a polar code of length $N=2^n$ and
dimension $K$. The source-bit vector is denoted by
$\mathbf{u}=(u_0,u_1,\ldots,u_{N-1})$. Let $\mathcal{I}$ and
$\mathcal{F}$ be the information set and the frozen set, respectively,
where $|\mathcal{I}|=K$ and $|\mathcal{F}|=N-K$. For each
$i\in\mathcal{F}$, the source bit is fixed to zero. The codeword is
generated over $\mathrm{GF}(2)$ as
\begin{equation*}
    \mathbf{c}=\mathbf{u}\mathbf{G}_N,
\end{equation*}
where
\begin{equation*}
    \mathbf{G}_N=\mathbf{F}^{\otimes n},\qquad
    \mathbf{F}=
    \genfrac{[}{]}{0pt}{}{1\quad 0}{1\quad 1}.
\end{equation*}

In the blind-recognition scenario considered in this paper, the code length
$N$ is assumed to be known at the receiver, whereas the dimension $K$ and
the frozen/information-set pattern are unknown. Suppose that $M$
independent codewords are transmitted using binary phase-shift keying
(BPSK) over an additive white Gaussian noise (AWGN) channel. The $i$-th
coded bit of the $m$-th codeword is mapped to
\begin{equation*}
    x_{m,i}=1-2c_{m,i}\in\{+1,-1\},
\end{equation*}
and the received sample is
\begin{equation*}
    r_{m,i}=x_{m,i}+n_{m,i},\qquad
    n_{m,i}\sim\mathcal{N}(0,\sigma^2),
\end{equation*}
where $0\le m\le M-1$ and $0\le i\le N-1$. The corresponding channel LLR is
\begin{equation}
    \lambda_{m,i}
    =
    \log
    \frac{p(r_{m,i}\mid c_{m,i}=0)}
         {p(r_{m,i}\mid c_{m,i}=1)}
    =
    \frac{2r_{m,i}}{\sigma^2}.
    \label{eq:channel_llr}
\end{equation}
Collecting the channel LLRs of all M received vectors yields
\begin{equation*}
    \boldsymbol{\Lambda}^{\mathrm{ch}}
    =
    \left[
    \begin{array}{cccc}
        \lambda_{0,0}     & \lambda_{0,1}     & \cdots & \lambda_{0,N-1} \\
        \lambda_{1,0}     & \lambda_{1,1}     & \cdots & \lambda_{1,N-1} \\
        \vdots            & \vdots            & \ddots & \vdots          \\
        \lambda_{M-1,0}   & \lambda_{M-1,1}   & \cdots & \lambda_{M-1,N-1}
    \end{array}
    \right],
\end{equation*}
which serves as the input for subsequent blind recognition.

\subsection{SC Recursion and Synthetic LLRs}

In SC decoding, the LLRs associated with the
source-bit positions are computed recursively on the polar decoding tree.
For a length-two polar transform with input LLRs $\lambda_0$ and
$\lambda_1$, the first source-bit LLR is obtained by the $f$-operation as
\begin{equation}
    \Lambda_0=\lambda_0\boxplus\lambda_1,
    \label{eq:sc_f_operation}
\end{equation}
where
\begin{equation*}
    a\boxplus b
    =
    2\operatorname{arctanh}
    \left(
        \tanh\frac{a}{2}
        \tanh\frac{b}{2}
    \right).
\end{equation*}
After the hard decision $\hat{u}_0$ is obtained from $\Lambda_0$, 
the second source-bit LLR is obtained by the $g$-operation as
\begin{equation}
    \Lambda_1=(-1)^{\hat{u}_0}\lambda_0+\lambda_1 .
    \label{eq:sc_g_operation}
\end{equation}

For a general polar code, the same two operations are applied recursively
to obtain the source-bit LLRs $\Lambda_0,\Lambda_1,\ldots,\Lambda_{N-1}$. When
the frozen and information set are known, the standard SC hard decision at
position $i$ is given by
\begin{equation}
    \hat{u}_i =
    \begin{cases}
        0, & i\in\mathcal{F},\\
        0, & i\in\mathcal{I},\ \Lambda_i\ge 0,\\
        1, & i\in\mathcal{I},\ \Lambda_i<0.
    \end{cases}
    \label{eq:sc_decision_rule}
\end{equation}

Let $W_i$ denote the $i$-th polar synthetic channel. When the corresponding
synthetic-channel output is $y$, its LLR is defined as
\begin{equation}
    \Lambda_i(y)
    =
    \log
    \frac{W_i(y\mid 0)}
         {W_i(y\mid 1)} .
    \label{eq:synthetic_llr_definition}
\end{equation}

Since the SC recursion at position $i$ depends on the previous decisions,
we introduce the SC-consistent event
\begin{equation}
    A_i=\{\hat{u}_0^{i-1}=u_0^{i-1}\},
    \label{eq:sc_consistent_event}
\end{equation}
where $u_0^{i-1}$ and $\hat{u}_0^{i-1}$ denote the true and estimated
source-bit prefixes, respectively, and $A_0$ is the sure event.

\subsection{BSCL Recognition Method}

The BSCL recognition method in [21] addresses the unknown information-set pattern problem through a list-search mechanism.
At each source-bit position, each surviving path is expanded into two candidate
branches, corresponding to the frozen-bit hypothesis and the information-bit hypothesis.

The local metrics used in BSCL are as follows: For the $l$-th
surviving path, let $\Lambda_{m,i}$ denote the synthetic LLR at
source-bit position $i$ for the $m$-th intercepted observation. The metric
increment under the frozen-bit hypothesis is
\begin{equation}
    \Delta^{(\mathcal{F},i)}
    =
    \frac{1}{M}
    \sum_{m=0}^{M-1}
    \log\left(1+e^{-\Lambda_{m,i}}\right),
    \label{eq:bscl_frozen_metric}
\end{equation}
whereas the metric increment under the information-bit hypothesis is
\begin{equation}
    \Delta^{(\mathcal{I},i)}
    =
    \log 2 .
    \label{eq:bscl_information_metric}
\end{equation}

To continue the SC recursion for the $l$-th surviving path, the temporary
decision under the frozen-bit hypothesis is fixed as
\begin{equation}
    \hat{u}_{m,i}^{(\mathcal{F})}
    =
    0,
    \label{eq:bscl_frozen_decision}
\end{equation}
whereas that under the information-bit hypothesis is
\begin{equation}
    \hat{u}_{m,i}^{(\mathcal{I})}
    =
    \begin{cases}
        0, & \Lambda_{m,i} \ge 0,\\
        1, & \Lambda_{m,i} < 0 .
    \end{cases}
    \label{eq:bscl_information_decision}
\end{equation}
These temporary decisions are used only to update the corresponding partial
sums for the subsequent $g$-operations.

For source-bit position $i$, each surviving path is expanded into two
candidate branches, whose accumulated metrics are given by
\begin{equation}
    \mathrm{PM}^{(i)}
    =
    \mathrm{PM}^{(i-1)}
    +
    \Delta^{(b,i)},
    \qquad
    b\in\{\mathcal{F},\mathcal{I}\}.
    \label{eq:bscl_metric_update}
\end{equation}
After sorting the candidate paths according to their accumulated metrics,
only the best $L_{\mathrm{list}}$ paths are retained. After all source-bit
positions have been processed, the path with the minimum accumulated metric is
selected, and its path labels form the estimated
information set $\hat{\mathcal{I}}$.

Note that the BSCL recognition method reduces to the BSC recognition when the list size is set to one. Under this condition, the list-based path extension is no longer performed. Instead, a single local decision is made at each source-bit position to determine the corresponding path label.

\subsection{Density Evolution for Polar Synthetic Channels}

Let $a_{\mathrm{ch}}$ denote the distribution of the channel LLR
conditioned on the transmitted coded bit $0$. For the BPSK-AWGN model
considered in this paper, we have
\begin{equation}
    a_{\mathrm{ch}}
    =
    \mathcal{N}\left(
        \frac{2}{\sigma^2},
        \frac{4}{\sigma^2}
    \right).
    \label{eq:initial_llr_density}
\end{equation}
The DE recursion is initialized by $a_0^{(0)}=a_{\mathrm{ch}}$.

At stage $t$, let $a_j^{(t)}$ denote the input-zero LLR distribution of
the $j$-th intermediate synthetic channel. The DE update follows the same
two operations as the SC recursion. For two independent random variables
$L_1$ and $L_2$ with distribution $a_j^{(t)}$, define
\begin{equation*}
    L^- = L_1\boxplus L_2,\qquad
    L^+ = L_1+L_2 .
\end{equation*}
The corresponding density recursion is
\begin{equation}
    a_{2j}^{(t+1)}
    =
    \mathcal{L}(L^-),
    \qquad
    a_{2j+1}^{(t+1)}
    =
    \mathcal{L}(L^+),
    \label{eq:de_density_recursion}
\end{equation}
where $\mathcal{L}(X)$ denotes the probability law of the random variable
$X$. Repeating \eqref{eq:de_density_recursion} for
$t=0,1,\ldots,n-1$ gives the final synthetic-channel LLR distributions
$a_0^{(n)},a_1^{(n)},\ldots,a_{N-1}^{(n)}$.

For the $i$-th synthetic channel, the input-zero LLR distribution is denoted by
\begin{equation}
    P_i^0
    =
    a_i^{(n)}
    =
    \mathcal{L}\left(\Lambda_i \mid u_i=0\right).
    \label{eq:Pi0_distribution}
\end{equation}
Similarly, the input-one LLR distribution is defined as
\begin{equation}
    P_i^1
    =
    \mathcal{L}\left(\Lambda_i \mid u_i=1\right).
    \label{eq:Pi1_distribution}
\end{equation}

The BPSK-AWGN channel considered in this paper is a binary-input
memoryless symmetric (BMS) channel, and the polar transform preserves
channel symmetry. Hence, the polar synthetic channels are also symmetric.
With the LLR definition in \eqref{eq:synthetic_llr_definition}, the
input-one and input-zero LLR distributions satisfy
\begin{equation}
    \frac{dP_i^1}{dP_i^0}(\ell)
    =
    e^{-\ell}.
    \label{eq:llr_symmetry_relation}
\end{equation}

\section{Simplified BSCL Recognition}

In this section, we first analyze the first recognition-error positions in the BSC recognition. It is observed that these positions are closely related to the corresponding upper-bound contribution terms in [22], which indicates that only a limited number of  bit positions play a dominant role in the recognition failure. Based on this observation, we propose a critical-set-aided simplified BSCL recognition method.

\subsection{First-Error Position Analysis}

\begin{table}[t]
\centering
\caption{First-Error Statistics Under Different Code Rates for $N=64$ and $M=500$}
\label{tab:first_error_position}
\footnotesize
\renewcommand{\arraystretch}{1.12}
\begin{tabular*}{\columnwidth}{@{\extracolsep{\fill}}cccccccc@{}}
\toprule
\multirow{2}{*}{$R$}
& \multirow{2}{*}{$E_s/N_0$ (dB)}
& \multirow{2}{*}{$P_{e}$}
& \multicolumn{5}{c}{First-error position (\%)} \\
\cmidrule(lr){4-8}
& & & $p_0$ & $p_1$ & $p_2$ & $p_4$ & $p_8$ \\
\midrule
\multirow{5}{*}{$1/4$}
& 0 & 0.88292 & 56.56  & 22.72 & 12.50 & 5.96 & 2.00 \\
& 1 & 0.63877 & 75.73  & 15.03 & 7.27  & 1.89 & 0.08 \\
& 2 & 0.40085 & 98.98  & 0.94  & 0.08  & 0.00 & 0.00 \\
& 3 & 0.12054 & 100.00 & 0.00  & 0.00  & 0.00 & 0.00 \\
& 4 & 0.00023 & 100.00 & 0.00  & 0.00  & 0.00 & 0.00 \\
\midrule
\multirow{5}{*}{$1/2$}
& 0 & 0.88016 & 56.43  & 22.89 & 12.39 & 5.93 & 2.10 \\
& 1 & 0.64053 & 75.51  & 15.18 & 7.25  & 1.97 & 0.08 \\
& 2 & 0.40158 & 98.98  & 0.96  & 0.06  & 0.00 & 0.00 \\
& 3 & 0.11973 & 100.00 & 0.00  & 0.00  & 0.00 & 0.00 \\
& 4 & 0.00022 & 100.00 & 0.00  & 0.00  & 0.00 & 0.00 \\
\midrule
\multirow{5}{*}{$3/4$}
& 0 & 0.89225 & 55.97  & 22.43 & 12.14 & 5.89 & 2.32 \\
& 1 & 0.63949 & 75.36  & 15.14 & 7.54  & 1.85 & 0.11 \\
& 2 & 0.40115 & 98.92  & 0.98  & 0.10  & 0.00 & 0.00 \\
& 3 & 0.12021 & 100.00 & 0.00  & 0.00  & 0.00 & 0.00 \\
& 4 & 0.00024 & 100.00 & 0.00  & 0.00  & 0.00 & 0.00 \\
\bottomrule
\end{tabular*}
\end{table}

\begin{table}[t]
\centering
\caption{Bhattacharyya-Parameter-Based Upper-Bound Error Contributions at Selected Positions for $N=64$ and $M=500$}
\label{tab:pe_upper_position}
\scriptsize
\renewcommand{\arraystretch}{1.12}
\setlength{\tabcolsep}{1.4pt}
\makebox[\columnwidth][l]{%
\begin{tabular*}{0.98\columnwidth}{@{\extracolsep{\fill}}cccccc@{}}
\toprule
\multirow{2}{*}{$E_s/N_0$ (dB)}
& \multicolumn{5}{c}{Upper-bound contribution} \\
\cmidrule(lr){2-6}
& $p_0$ & $p_1$ & $p_2$ & $p_4$ & $p_8$ \\
\midrule
0 & $1.0{\times}10^{0}$ & $9.9{\times}10^{-1}$ & $9.8{\times}10^{-1}$ & $9.5{\times}10^{-1}$ & $8.2{\times}10^{-1}$ \\
1 & $1.0{\times}10^{0}$ & $8.9{\times}10^{-1}$ & $7.9{\times}10^{-1}$ & $5.7{\times}10^{-1}$ & $2.3{\times}10^{-1}$ \\
2 & $9.9{\times}10^{-1}$ & $3.6{\times}10^{-1}$ & $1.7{\times}10^{-1}$ & $3.3{\times}10^{-2}$ & $9.4{\times}10^{-4}$ \\
3 & $8.0{\times}10^{-1}$ & $4.7{\times}10^{-3}$ & $2.2{\times}10^{-4}$ & $8.2{\times}10^{-7}$ & $4.4{\times}10^{-11}$ \\
4 & $1.2{\times}10^{-1}$ & $1.2{\times}10^{-9}$ & $1.5{\times}10^{-13}$ & $1.6{\times}10^{-19}$ & $6.1{\times}10^{-28}$ \\
\bottomrule
\end{tabular*}%
}
\end{table}

In  blind recognition scenario, the recognition result is success only when all source-bit positions are correctly recognized.
Therefore,  the first-error position is especially important in a failed trial, because it directly affect the subsequent SC recursion
through the $g$-operations.

To further analyze the origin of the  recognition failure event, Table I reports the first-error statistics obtained from failed trials
for $N=64$ and $M=500$ under BSC recognition.  Here, $P_{e}$ denotes the  recognition error
probability, and $p_i$ is the proportion of failed trials whose first error occurs at position $i$. Only the dominant
positions $0,1,2,4,$ and $8$ are listed for compactness.

It can be observed that the first-error positions are far from uniformly distributed. For all three code rates, most first errors occur at
positions $0,1,2,4,$ and $8$, and the distributions under $R=1/4$, $R=1/2$, and $R=3/4$ are very similar. This suggests that the dominant
first-error positions are mainly determined by the reliability of the corresponding polar synthetic channels.
As $E_s/N_0$ increases, the first-error events become increasingly dominated by position $0$. For example, when $E_s/N_0=2$ dB, 
the first error occurs at position $0$ in nearly $99\%$ of the failed trials.
When $E_s/N_0\geq 3$ dB, all observed first errors occur at this position.

To further understand this phenomenon, Table II lists the position-wise upper-bound contribution terms of the same positions according to the
Bhattacharyya-parameter-based analysis in [22]. 
These terms quantify the relative difficulty of distinguishing the frozen and information hypotheses at the corresponding positions.

The comparison between Table I and Table II shows a similar trend between the empirical first-error distribution and the upper-bound
contribution terms. At low SNRs, positions $0,1,2,4,$ and $8$ all have large contribution terms. For example, at $E_s/N_0=0$ dB, the
contribution terms of these positions range from $8.2\times 10^{-1}$ to $1.0$. Correspondingly, the first errors in Table I are distributed over
these positions, where position $0$ accounts for about $56\%$ of the failed trials, and positions $1$, $2$, $4$, and $8$ also have noticeable
percentages. As the SNR increases, the contribution terms of positions $1,2,4,$ and $8$ decrease rapidly, whereas position $0$ remains the
dominant one. For instance, at $E_s/N_0=2$ dB, the contribution term of position $8$ drops to $9.4\times 10^{-4}$, and its first-error
percentage in Table I becomes zero. In contrast, position $0$ still has a large contribution term, and its first-error percentage is close to
$99\%$. When the SNR further increases, the contribution terms of the listed positions except position $0$ become almost negligible, and the
observed first errors are also concentrated at position $0$. This similarity suggests that the position-wise contribution terms in [22]
can provide a useful reference for locating the positions that are likely to trigger early recognition errors.

\subsection{Critical-Set Selection and SBSCL Procedure}

In BSCL, the two-hypothesis path expansion is used to protect the recognition process from early recognition errors. 
However, the first-error analysis in the preceding subsection shows that such early
errors are not uniformly distributed over all source-bit positions. Instead, they are concentrated on a small number of positions, and these
positions are closely related to the large position-wise contribution terms. Therefore, it is unnecessary to apply the full BSCL path expansion at
every position. 
The expansion should be retained only at the positions that are likely to trigger early recognition errors, and these positions are referred to as the critical set.

For source-bit position $i$, let $U_i$ denote its contribution to the upper bound for a given SNR, code length $N$, and number
of intercepted observations $M$. Note that the parameter $U_i$ can be computed offline, since it depends only on the above condition.
A larger $U_i$ indicates that the frozen-bit and information-bit hypotheses at position $i$ are more difficult to distinguish, and
therefore this position is more likely to cause an recognition error.
Based on this contribution term, the critical set is defined as
\begin{equation}
    \mathcal{J}_{\eta}
    =
    \left\{
        i:U_i>\eta,\ 0\le i<N
    \right\},
    \label{eq:expansion_set}
\end{equation}
where $\eta$ is a preset threshold. 

The positions in $\mathcal{J}_{\eta}$ are processed using the same two-hypothesis path expansion as in BSCL. For positions outside
$\mathcal{J}_{\eta}$, the path expansion operation is suppressed. 
The threshold $\eta$ controls the size of critical set. A smaller $\eta$ includes more positions in
$\mathcal{J}_{\eta}$ and makes SBSCL closer to BSCL, whereas a larger $\eta$ includes fewer positions in $\mathcal{J}_{\eta}$. 
In particular, when the critical set is empty, all list-related operations are eliminated, which reduces the computational complexity associated with path expansion, path-state copying, accumulated path-metric updates, candidate sorting, and pruning.

SBSCL uses the same local metrics and temporary decisions as BSCL. For the $l$-th surviving path at source-bit position $i$, the metric
increments under the frozen-bit and information-bit hypotheses are computed according to \eqref{eq:bscl_frozen_metric} and
\eqref{eq:bscl_information_metric}, respectively. The corresponding temporary decisions are obtained according to
\eqref{eq:bscl_frozen_decision} and \eqref{eq:bscl_information_decision}. 
For a source-bit position $i\in\mathcal{J}_{\eta}$, SBSCL follows the same branching rule as BSCL. Each surviving path is expanded into two
candidate branches, corresponding to the frozen-bit and information-bit hypotheses. The accumulated metrics of the two branches are updated by
\eqref{eq:bscl_metric_update}. The generated candidate branches are then sorted in ascending order of accumulated path metric, and only the best
$L_{\rm list}$ paths are retained.
For a source-bit position $i\notin\mathcal{J}_{\eta}$, SBSCL does not split the surviving paths. Instead,
each surviving path performs the BSC recognition, i.e., each path is updated by
comparing the two local metric increments. If
\begin{equation*}
    \Delta^{(\mathcal{F},i)}
    \le
    \Delta^{(\mathcal{I},i)} ,
\end{equation*}
the $i$-th position is labeled as $\mathcal{F}$; otherwise, it is labeled as $\mathcal{I}$. The accumulated path metric is updated as
\begin{equation}
    PM^{(i)}
    =
    PM^{(i-1)}
    +
    \min\left\{
        \Delta^{(\mathcal{F},i)},
        \Delta^{(\mathcal{I},i)}
    \right\}.
    \label{eq:sbscl_single_metric}
\end{equation}
The recursive LLR state and partial sums are then updated according to the selected label.

The overall SBSCL procedure is summarized in Algorithm~\ref{alg:sbscl}. Lines 3--4 initialize the path metric, path
labels, recursive LLR states, and partial sums. Lines 5--35 process the source-bit positions sequentially. If the current position belongs to
the critical set, lines 6--20 follow the same list operation as BSCL, including two-hypothesis path expansion, path-state copying, candidate sorting, and
pruning. If the current position is outside the critical set, lines 21--34 update each surviving path only once according to the local metric comparison.
Finally, lines 36--37 select the surviving path with the minimum accumulated path metric and recover the estimated information set.

\begin{algorithm}[!t]
\caption{SBSCL Algorithm}
\label{alg:sbscl}
\begin{algorithmic}[1]
\STATE \textbf{Input:} Channel LLR matrix $\Lambda^{\rm ch}$, code length
$N$, list size $L_{\rm list}$, and critical set $\mathcal{J}_{\eta}$
\STATE \textbf{Output:} Estimated information set $\hat{\mathcal{I}}$

\STATE Activate one initial path with zero metric
\STATE Initialize path labels, recursive LLR states, and partial sums

\FOR{$i=0$ to $N-1$}
    \IF{$i\in\mathcal{J}_{\eta}$}
        \FOR{each surviving path}
            \FOR{$m=0$ to $M-1$}
                \STATE Compute the decision LLR $\Lambda_{m,i}$ via the $f$- and $g$-operations
                \STATE Set $\hat{u}_{m,i}^{(\mathcal{F})}$ according to \eqref{eq:bscl_frozen_decision}
                \STATE Generate $\hat{u}_{m,i}^{(\mathcal{I})}$ according to \eqref{eq:bscl_information_decision}
            \ENDFOR  
            \STATE Generate candidate branches under the frozen-bit hypothesis and the information-bit hypothesis
            \STATE  Compute $\Delta^{(\mathcal{F},i)}$ according to \eqref{eq:bscl_frozen_metric}
            \STATE  Set $\Delta^{(\mathcal{I},i)}$ according to \eqref{eq:bscl_information_metric} 
            \STATE Update accumulated path metrics via \eqref{eq:bscl_metric_update}
        \ENDFOR
        \STATE Sort all candidate branches in ascending order of accumulated path metric 
        \STATE Retain the best $L_{\mathrm{list}}$ branches
        \STATE Update the corresponding path labels, recursive LLR states, and partial sums

    \ELSE
        \FOR{each surviving path}
            \FOR{$m=0$ to $M-1$}
                \STATE Compute the decision LLR $\Lambda_{m,i}$ via the $f$- and $g$-operations
                \STATE Set $\hat{u}_{m,i}^{(\mathcal{F})}$ according to \eqref{eq:bscl_frozen_decision}
                \STATE Generate $\hat{u}_{m,i}^{(\mathcal{I})}$ according to \eqref{eq:bscl_information_decision}
            \ENDFOR 
            \STATE  Compute $\Delta^{(\mathcal{F},i)}$ according to \eqref{eq:bscl_frozen_metric}
            \STATE  Set $\Delta^{(\mathcal{I},i)}$ according to \eqref{eq:bscl_information_metric}
            \STATE  Compare $\Delta^{(\mathcal{F},i)}$ and $\Delta^{(\mathcal{I},i)}$, and select the one with the smaller metric.
            \STATE Update accumulated path metrics via \eqref{eq:sbscl_single_metric}
            \STATE Update the corresponding path labels, recursive LLR states, and partial sums
        \ENDFOR
    \ENDIF
\ENDFOR

\STATE Select the surviving path with the minimum accumulated path metric
\STATE Recover $\hat{\mathcal{I}}$ from its path-label sequence
\end{algorithmic}
\end{algorithm}

\begin{figure}[t]
  \centering
  \includegraphics[width=1\linewidth]{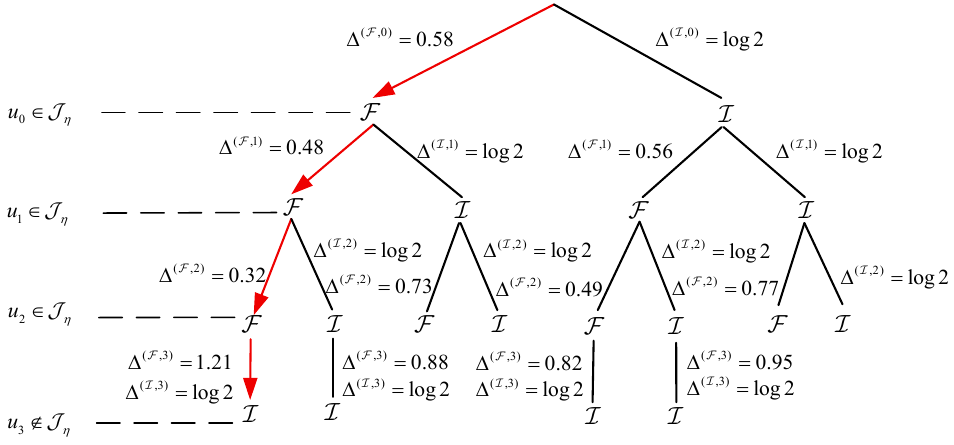}
  \caption{SBSCL Recognition Decoding Tree  with $N=4$ and $L_{list}=4$}
  \label{fig:Decoding Tree}
\end{figure}

Fig. 1 illustrates the recognition decoding tree of the SBSCL algorithm with $N=4$ and $L_{list}=4$. It can be observed that the bit positions with indices 0, 1, and 2 are  included in the critical set $\mathcal{J}_{\eta}$. Therefore, the corresponding paths are expanded under two hypotheses.  At the positions $u_0$ and $u_1$, the number of candidate paths does not exceed the list size, and thus no pruning operation is performed. After the expansion at position $u_2$, the number of candidate paths increases to 8. Then, only the 4 paths with the smallest accumulated path metrics are retained. Since $u_3$ does not belong to the critical set $\mathcal{J}_{\eta}$, each surviving path is extended according to a single-path decision. Finally, the path indicated by the red arrows is selected as the recognition result because it has the minimum accumulated path metric.

\section{Density-Evolution Bounds for Blind Recognition}

In this section, to improve the reliability of critical-set selection and provide a more accurate performance analysis, we derive a DE-based bound for the blind recognition  under the SC-consistent condition. In the proposed DE-bound, the frozen-bit and information-bit distributions are characterized by the DE-evolved synthetic-LLR distributions. Based on these distributions, the position-type overlap coefficient is directly obtained, while the Chernoff parameter is optimized to tighten the resulting upper bound.

\subsection{Position-Type Test from DE Method}

For the $i$-th source-bit position, set
\[
    P_i \triangleq P_i^0 ,
\]
where $P_i^0$ denotes the DE-evolved input-zero LLR distribution in
\eqref{eq:Pi0_distribution}. Under the frozen-bit hypothesis, the
source bit is fixed to zero. Under the information-bit hypothesis, the
source bit is modeled as equiprobable. Hence, the two hypotheses induce
the LLR distributions
\begin{equation}
\left\{
\begin{aligned}
H_{F,i} &: \Lambda_i \sim P_i,\\
H_{I,i} &: \Lambda_i \sim Q_i .
\end{aligned}
\right.
\label{eq:position_type_test}
\end{equation}
For the information-bit hypothesis, the corresponding distribution is
\begin{equation}
    Q_i
    =
    \frac{1}{2}P_i^0+\frac{1}{2}P_i^1 .
    \label{eq:Qi_definition}
\end{equation}
Using the symmetry relation in \eqref{eq:llr_symmetry_relation}, we obtain
\begin{equation}
    \frac{dQ_i}{dP_i}(\ell)
    =
    \frac{1+e^{-\ell}}{2}.
    \label{eq:Qi_density_ratio}
\end{equation}
Therefore, the log-likelihood ratio for discriminating the information model from the frozen model is
\begin{equation}
    \Gamma_i(\ell)
    =
    \log\frac{dQ_i}{dP_i}(\ell)
    =
    \log(1+e^{-\ell})-\log 2 .
    \label{eq:type_llr}
\end{equation}

Define the single-observation recognition metric as
\begin{equation}
    C_i(\ell)=\log(1+e^{-\ell}) .
    \label{eq:soft_metric}
\end{equation}
Then
\begin{equation}
    C_i(\ell)
    =
    \Gamma_i(\ell)+\log 2 .
    \label{eq:metric_shift_relation}
\end{equation}
This is the same shifted-LLR relation as in [22]. While
[22] derives it from the synthetic-channel output distributions,
we obtain the same relation directly from the DE-evolved LLR
distributions. The neutral threshold is therefore still $\log 2$.

For the $m$-th intercepted observation, let $\Lambda_{m,i}$ denote the
SC-consistent synthetic LLR at source-bit position $i$. With $M$
independent intercepted observations, the accumulated recognition metric and the
corresponding accumulated log-likelihood ratio are defined as
\begin{equation}
    S_i^{(M)}
    =
    \sum_{m=0}^{M-1} C_i(\Lambda_{m,i}),
    \qquad
    G_i^{(M)}
    =
    \sum_{m=0}^{M-1} \Gamma_i(\Lambda_{m,i}) .
    \label{eq:accumulated_statistics}
\end{equation}
From \eqref{eq:metric_shift_relation}, these two statistics satisfy
\begin{equation}
    S_i^{(M)}
    =
    G_i^{(M)}+M\log 2 .
    \label{eq:S_G_relation}
\end{equation}
Under equal local priors and equal misclassification costs, the neutral
position-wise likelihood-ratio test is
\begin{equation}
    G_i^{(M)}
    \mathop{\gtrless}_{H_{F,i}}^{H_{I,i}}
    0 .
    \label{eq:neutral_test_G}
\end{equation}
By \eqref{eq:S_G_relation}, this is equivalent to comparing the averaged
recognition metric with $\log 2$, i.e.,
\begin{equation}
    \frac{1}{M}S_i^{(M)}
    \mathop{\gtrless}_{H_{F,i}}^{H_{I,i}}
    \log 2 .
    \label{eq:neutral_test_S}
\end{equation}

\subsection{DE-Based Upper and Lower Bounds}

We now derive DE-based bounds for the neutral position-wise likelihood-ratio test in \eqref{eq:neutral_test_G}. Under the
SC-consistent condition, the test at position $i$ distinguishes the frozen-bit distribution $P_i$ from the information-bit
distribution $Q_i$.

Define the two one-sided error probabilities as
\begin{align}
P_{F\to I,i}^{(M)}
&\triangleq
\Pr
\left\{
G_i^{(M)}>0
\mid H_{F,i},A_i
\right\},
\label{eq:PFI_def} \\
P_{I\to F,i}^{(M)}
&\triangleq
\Pr
\left\{
G_i^{(M)}\le 0
\mid H_{I,i},A_i
\right\}.
\label{eq:PIF_def}
\end{align}
Let $\mathcal{E}_i$ denote the local position-type error event. With equal local priors and equal misclassification costs,
\begin{equation}
P_{e,i}^{(M)}
\triangleq
\Pr(\mathcal{E}_i\mid A_i)
=
\frac{1}{2}P_{F\to I,i}^{(M)}
+
\frac{1}{2}P_{I\to F,i}^{(M)} .
\label{eq:local_error}
\end{equation}

For two probability measures $P$ and $Q$, let
\[
B(P,Q)
\triangleq
\int \sqrt{dP dQ}
\]
denote the Bhattacharyya coefficient. For the position-type test at index $i$, define the overlap coefficient
\begin{equation}
\Omega_i
\triangleq
B(P_i,Q_i)
=
\int \sqrt{dP_i dQ_i}.
\label{eq:omega_def}
\end{equation}
Using  \eqref{eq:Qi_density_ratio},
$\Omega_i$ can be computed in the LLR domain as
\begin{equation}
\Omega_i
=
\mathbb{E}_{L_i\sim P_i}
\left[
\sqrt{
\frac{1+e^{-L_i}}{2}
}
\right].
\label{eq:omega_llr}
\end{equation}

For the upper bound, we use a Chernoff-type coefficient. For
$0\le s\le 1$, define
\begin{equation}
\mathcal{T}_i(s)
\triangleq
\int
(dP_i)^{1-s}
(dQ_i)^s .
\label{eq:tilted_overlap_def}
\end{equation}
Equivalently,
\[
\mathcal{T}_i(s)
=
\mathbb{E}_{L_i\sim P_i}
\left[
e^{s\Gamma_i(L_i)}
\right]
=
\mathbb{E}_{L_i\sim Q_i}
\left[
e^{-(1-s)\Gamma_i(L_i)}
\right].
\]
The DE-based Chernoff coefficient is defined as
\begin{equation}
\beta_i
\triangleq
\inf_{0\le s\le 1}
\mathcal{T}_i(s).
\label{eq:beta_def}
\end{equation}
Since $\Omega_i=\mathcal{T}_i(1/2)$, we have
\[
\beta_i\le \Omega_i .
\]
Using \eqref{eq:Qi_density_ratio}, $\beta_i$ can be evaluated directly from the DE-evolved input-zero LLR distribution as
\begin{equation}
\beta_i
=
\inf_{0\le s\le 1}
\mathbb{E}_{L_i\sim P_i}
\left[
\left(
\frac{1+e^{-L_i}}{2}
\right)^s
\right].
\label{eq:beta_llr}
\end{equation}

\textbf{Theorem 1:}
Under $A_i$ and the neutral decision rule,
\begin{equation}
P_{F\to I,i}^{(M)}
\le
\beta_i^M,
\qquad
P_{I\to F,i}^{(M)}
\le
\beta_i^M .
\label{eq:one_sided_bound}
\end{equation}
Moreover,
\begin{equation}
\frac{1}{2}
\left(
1-\sqrt{1-\Omega_i^{2M}}
\right)
\le
P_{e,i}^{(M)}
\le
\beta_i^M .
\label{eq:position_bound}
\end{equation}

\textit{Proof:}
Conditioned on $H_{F,i}$ and $A_i$, the $M$ synthetic LLRs are independent with common distribution $P_i$. 
For any $0<s\le 1$, Markov's inequality gives
\begin{align*}
P_{F\to I,i}^{(M)}
&=
\Pr
\left\{
G_i^{(M)}>0
\mid H_{F,i},A_i
\right\} \\
&=
\Pr
\left\{
e^{sG_i^{(M)}}>1
\mid H_{F,i},A_i
\right\} \\
&\le
\mathbb{E}
\left[
e^{sG_i^{(M)}}
\mid H_{F,i},A_i
\right] \\
&=
\left(
\mathbb{E}_{L_i\sim P_i}
\left[
e^{s\Gamma_i(L_i)}
\right]
\right)^M \\
&=
\mathcal{T}_i(s)^M .
\end{align*}
Taking the infimum over $0<s\le1$ and using the continuity
of $\mathcal{T}_i(s)$ on $[0,1]$ gives
\[
P_{F\to I,i}^{(M)}
\le
\beta_i^M .
\]

Similarly, conditioned on $H_{I,i}$ and $A_i$, the $M$ synthetic LLRs are independent with common distribution
$Q_i$. For any $0\le s<1$, Markov's inequality gives
\begin{align*}
P_{I\to F,i}^{(M)}
&=
\Pr
\left\{
G_i^{(M)}\le 0
\mid H_{I,i},A_i
\right\} \\
&=
\Pr
\left\{
e^{-(1-s)G_i^{(M)}}\ge 1
\mid H_{I,i},A_i
\right\} \\
&\le
\mathbb{E}
\left[
e^{-(1-s)G_i^{(M)}}
\mid H_{I,i},A_i
\right] \\
&=
\left(
\mathbb{E}_{L_i\sim Q_i}
\left[
e^{-(1-s)\Gamma_i(L_i)}
\right]
\right)^M \\
&=
\mathcal{T}_i(s)^M .
\end{align*}
Taking the infimum over $0\le s<1$ and using the continuity
of $\mathcal{T}_i(s)$ on $[0,1]$ yields
\[
P_{I\to F,i}^{(M)}
\le
\beta_i^M .
\]
Together with \eqref{eq:local_error}, the two one-sided bounds prove the upper bound in \eqref{eq:position_bound}.

For the lower bound, we use the standard equal-prior binary testing identity and the Bhattacharyya-TV inequality [23].
Let \(P_i^{\bullet M}\) and \(Q_i^{\bullet M}\) denote the \(M\)-fold product measures induced by \(P_i\) and \(Q_i\), respectively. Equivalently, for \(M\) observed samples \(y_1,\ldots,y_M\), the corresponding joint densities are given by
\[
\begin{aligned}
P_i^{\bullet M}(y_1,\ldots,y_M)
&=
\prod_{m=1}^{M} P_i(y_m), \\
Q_i^{\bullet M}(y_1,\ldots,y_M)
&=
\prod_{m=1}^{M} Q_i(y_m).
\end{aligned}
\]
Then
\[
P_{e,i}^{(M)}
=
\frac{1}{2}
\left(
1-
\left\|
P_i^{\bullet M}
-
Q_i^{\bullet M}
\right\|_{\rm TV}
\right).
\]
Using
\[
\|P-Q\|_{\rm TV}
\le
\sqrt{1-B(P,Q)^2}
\]
and
\[
B(P_i^{\bullet M},Q_i^{\bullet M})
=
B(P_i,Q_i)^M
=
\Omega_i^M,
\]
we obtain
\[
P_{e,i}^{(M)}
\ge
\frac{1}{2}
\left(
1-\sqrt{1-\Omega_i^{2M}}
\right).
\]
This proves the lower bound in \eqref{eq:position_bound}.
\hfill $\square$

In numerical DE, if $P_i$ is represented by grid points $\ell_k$ and masses $w_{i,k}$, the objective in
\eqref{eq:beta_llr} is evaluated as
\[
\sum_k
w_{i,k}
\left(
\frac{1+e^{-\ell_k}}{2}
\right)^s .
\]
Since $\log \mathcal{T}_i(s)$ is convex in $s$, $\beta_i$ is obtained by a one-dimensional search over $[0,1]$.

Compared with the Bhattacharyya-parameter-based bound in [22], the proposed coefficients are evaluated directly from
the DE-evolved LLR distribution. In particular, $\Omega_i=\mathcal{T}_i(1/2)$ is used in the lower bound,
whereas the optimized Chernoff coefficient $\beta_i=\inf_{0\le s\le1}\mathcal{T}_i(s)$ is used in the upper bound.

\textbf{Corollary 1:}
Under the ideal SC-consistent sequence-level model, let $\mathcal{E}_{\rm tot}$ denote
the event that at least one source-bit position is incorrectly recognized in the neutral position-type testing process.
Then
\begin{equation}
\Pr(\mathcal{E}_{\rm tot})
\le
\sum_{i=0}^{N-1}
\beta_i^M .
\label{eq:seq_upper}
\end{equation}
Moreover,
\begin{equation}
\Pr(\mathcal{E}_{\rm tot})
\ge
1-
\prod_{i=0}^{N-1}
\left[
1-
\frac{1}{2}
\left(
1-\sqrt{1-\Omega_i^{2M}}
\right)
\right].
\label{eq:seq_lower}
\end{equation}

\textit{Proof:}
Define the recognition-consistent prefix event
\begin{equation}
R_i
=
\bigcap_{j=0}^{i-1}
\mathcal{E}_j^c,
\qquad
i=0,1,\ldots,N-1,
\label{eq:prefix_event}
\end{equation}
where the empty intersection for $i=0$ is the sure event. In the ideal SC-consistent sequence-level model, conditioning
on $R_i$ means that all previous position-type decisions are correct and that the $i$-th local test is evaluated with event $A_i$.

Let
\[
D_i=R_i\cap \mathcal{E}_i .
\]
Then $D_i$ is the event that the first recognition error occurs at position $i$, and
\[
\mathcal{E}_{\rm tot}
=
\bigcup_{i=0}^{N-1}
D_i .
\]
The events $D_0,D_1,\ldots,D_{N-1}$ are disjoint. Hence,
\begin{align*}
\Pr(\mathcal{E}_{\rm tot})
&=
\sum_{i=0}^{N-1}
\Pr(D_i) \\
&=
\sum_{i=0}^{N-1}
\Pr(R_i)\Pr(\mathcal{E}_i\mid R_i).
\end{align*}
Under the ideal SC-consistent model, conditioning on $R_i$ ensures that the $i$-th decision is evaluated with the correct
SC prefix. Hence, the position-wise bound in \eqref{eq:one_sided_bound} applies, and
\[
\Pr(\mathcal{E}_i\mid R_i)
\le
\beta_i^M .
\]
It follows that
\[
\Pr(\mathcal{E}_{\rm tot})
\le
\sum_{i=0}^{N-1}
\Pr(R_i)\beta_i^M
\le
\sum_{i=0}^{N-1}
\beta_i^M ,
\]
which proves \eqref{eq:seq_upper}.

\begin{figure}[t]
  \centering
  \includegraphics[width=1\linewidth]{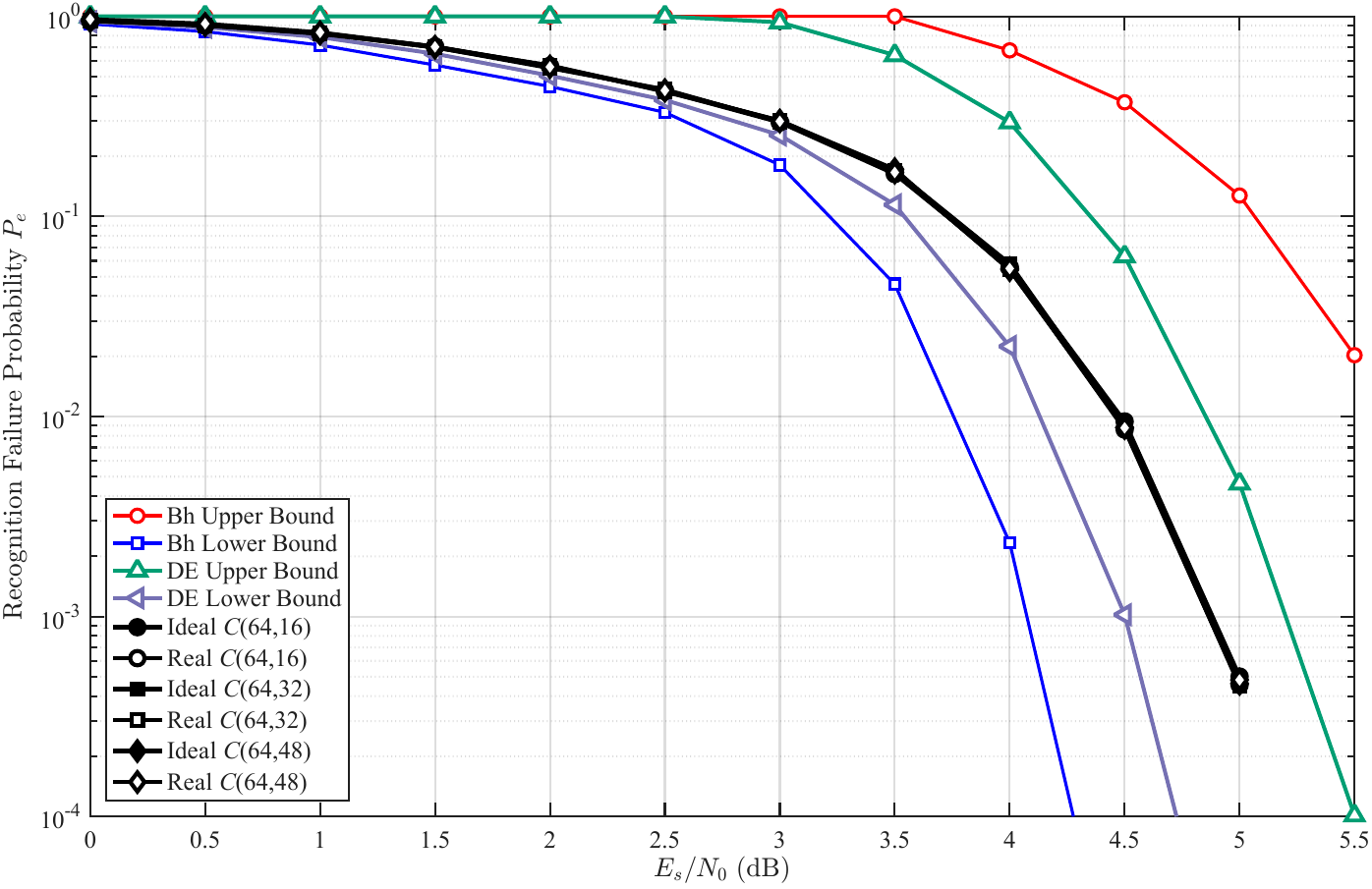}
  \caption{Performance comparison with $N=64,M=100$.}
  \label{fig:64M100}
\end{figure}

\begin{figure}[t]
  \centering
  \includegraphics[width=1\linewidth]{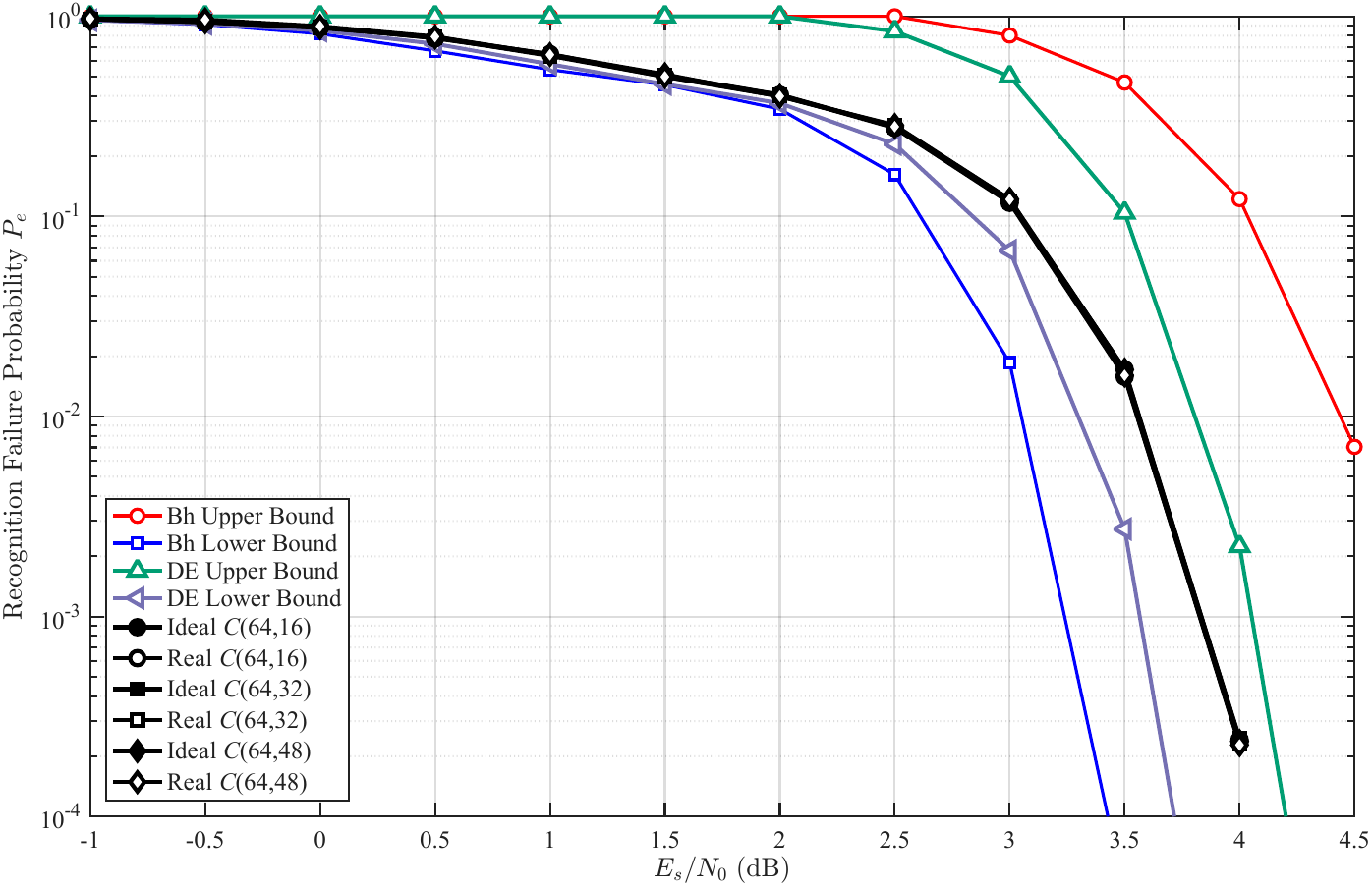}
  \caption{Performance comparison with $N=64,M=500$.}
  \label{fig:64M500}
\end{figure}

\begin{figure}[t]
  \centering
  \includegraphics[width=1\linewidth]{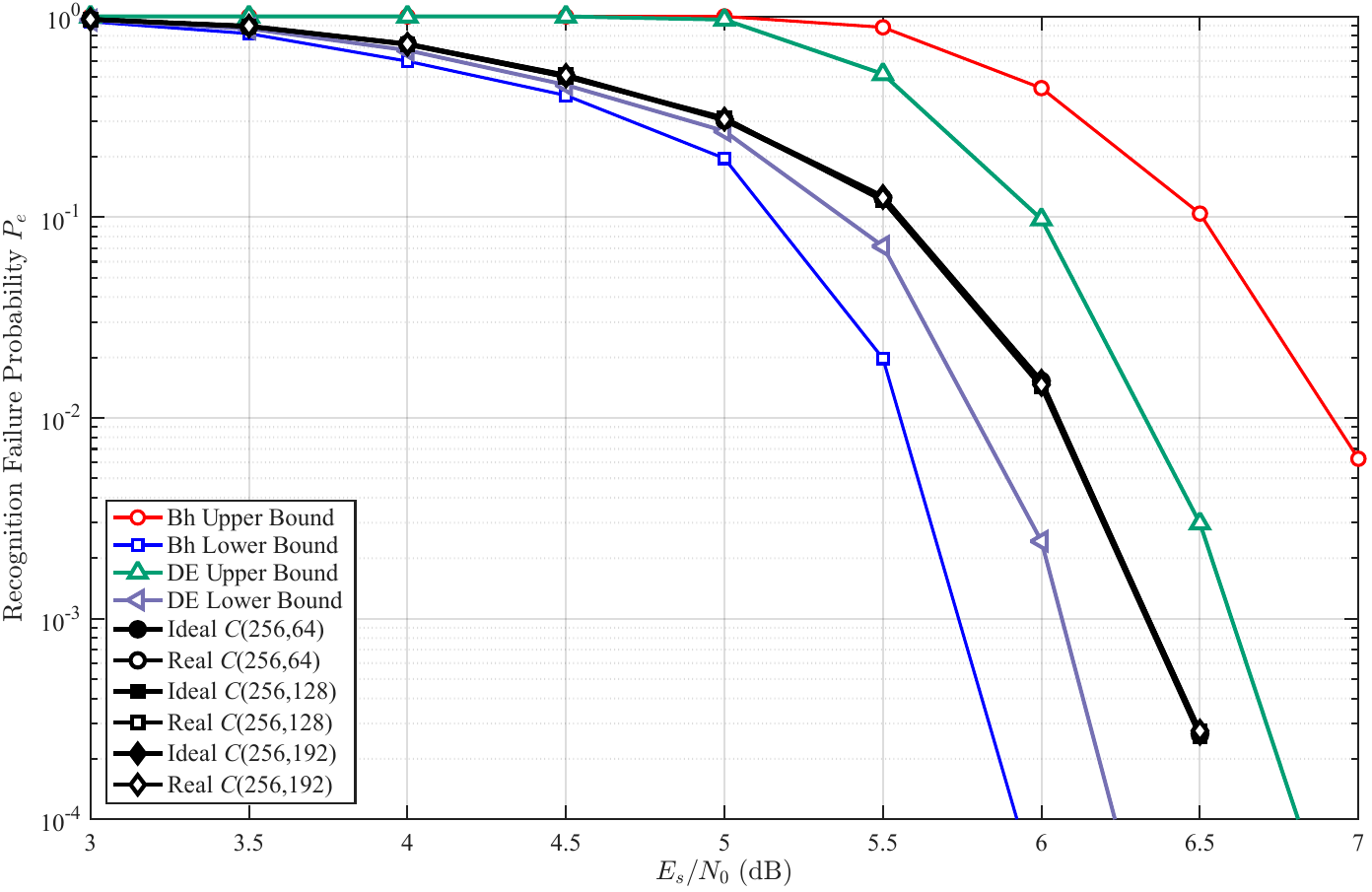}
  \caption{Performance comparison with $N=256,M=100$.}
  \label{fig:256M100}
\end{figure}

\begin{figure}[t]
  \centering
  \includegraphics[width=1\linewidth]{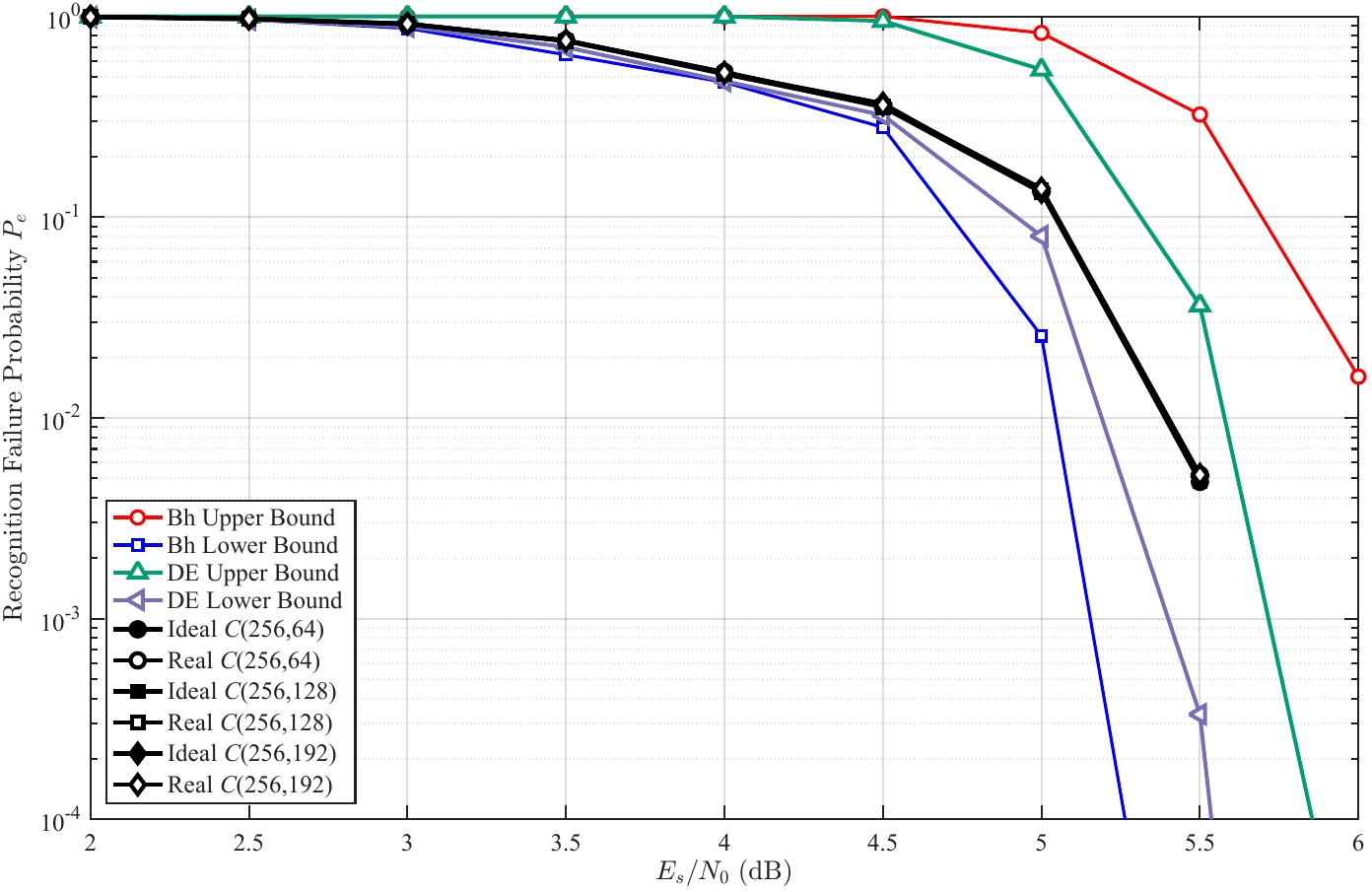}
  \caption{Performance comparison with $N=256,M=500$.}
  \label{fig:256M500}
\end{figure}

\begin{figure}[t]
  \centering
  \includegraphics[width=1\linewidth]{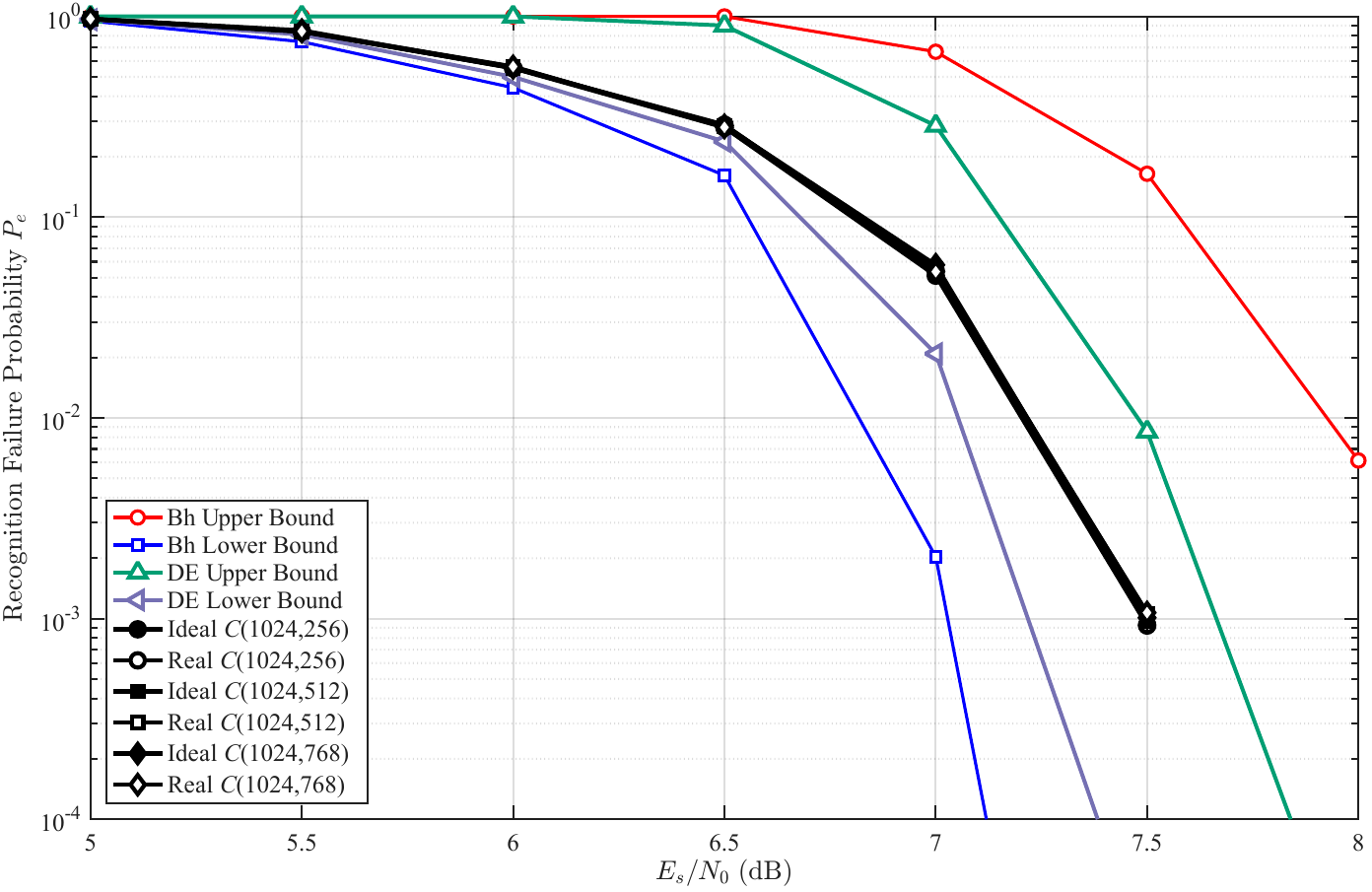}
  \caption{Performance comparison with $N=1024,M=100$.}
  \label{fig:1024M100}
\end{figure}

\begin{figure}[t]
  \centering
  \includegraphics[width=1\linewidth]{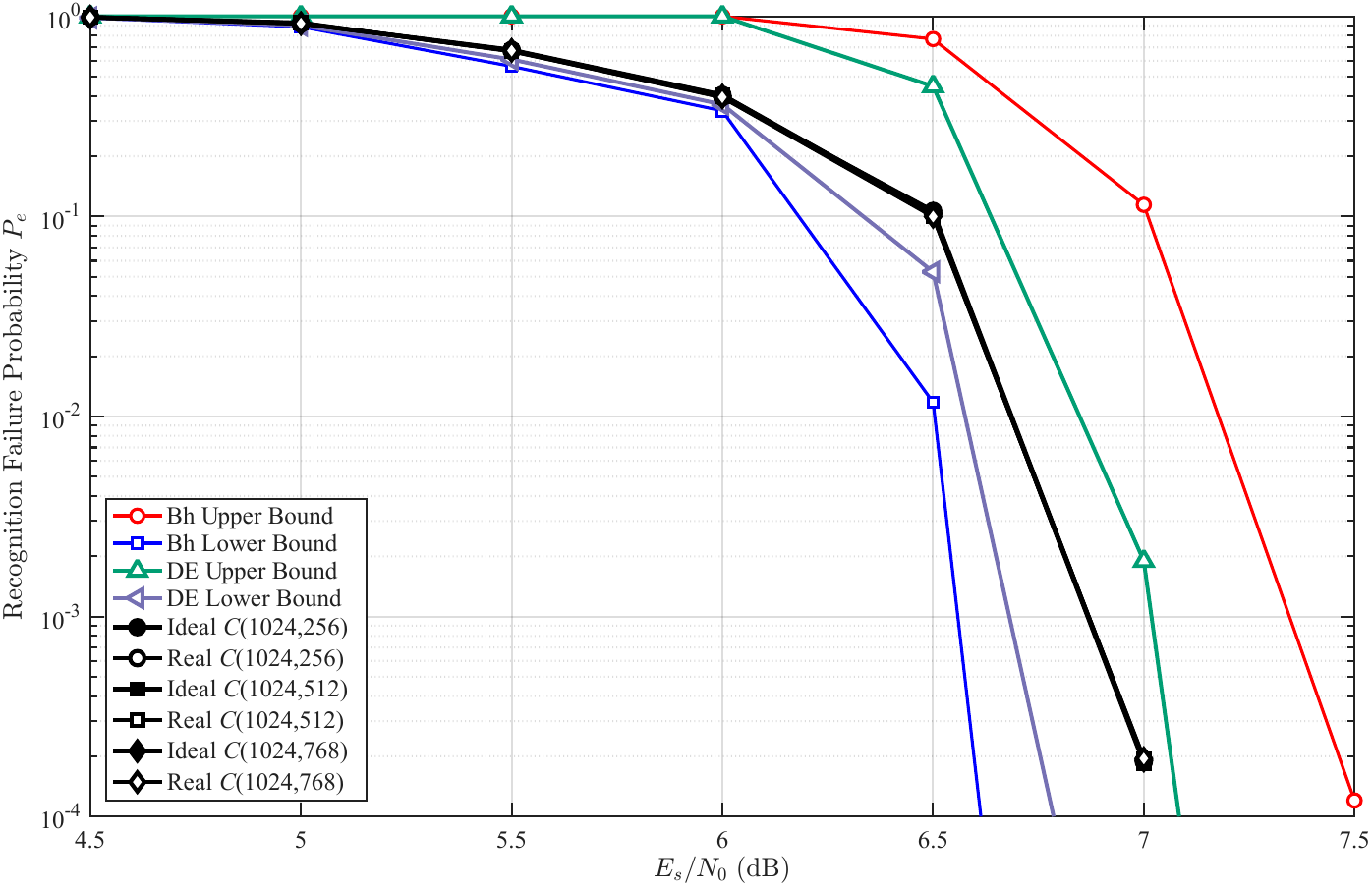}
  \caption{Performance comparison with $N=1024,M=500$.}
  \label{fig:1024M500}
\end{figure}

We next prove the sequence-level lower bound. Since no sequence error occurs if and only if all local position-type decisions are correct,
\[
\mathcal{E}_{\rm tot}^c
=
\bigcap_{i=0}^{N-1}
\mathcal{E}_i^c .
\]
Using the chain rule with the events $R_i$, we have
\begin{align*}
\Pr(\mathcal{E}_{\rm tot}^c)
&=
\prod_{i=0}^{N-1}
\Pr(\mathcal{E}_i^c\mid R_i) \\
&=
\prod_{i=0}^{N-1}
\left(
1-\Pr(\mathcal{E}_i\mid R_i)
\right).
\end{align*}
Equivalently,
\begin{equation}
\Pr(\mathcal{E}_{\rm tot})
=
1-
\prod_{i=0}^{N-1}
\left(
1-\Pr(\mathcal{E}_i\mid R_i)
\right).
\label{eq:seq_chain}
\end{equation}
Under the same neutral equal-prior formulation, \eqref{eq:position_bound} applies to the local test conditioned
on $R_i$, and hence
\[
\Pr(\mathcal{E}_i\mid R_i)
\ge
\frac{1}{2}
\left(
1-\sqrt{1-\Omega_i^{2M}}
\right).
\]
Since the function
\[
1-\prod_{i=0}^{N-1}(1-x_i)
\]
is monotonically increasing in each $x_i\in[0,1]$, substituting the above position-wise lower bound into \eqref{eq:seq_chain}
yields \eqref{eq:seq_lower}. 
\hfill $\square$

\section{Simulation Results}

This section presents Monte Carlo simulations to evaluate the proposed DE-based bounds and the SBSCL recognition method. The proposed bounds are first compared with the Bhattacharyya-parameter-based bounds in [22]. Then, the recognition performance and complexity reduction of SBSCL are evaluated.
In this section, all polar codes are constructed by the Gaussian-approximation (GA) method [24] at $E_b/N_0=2$ dB. The coded bits are BPSK modulated and transmitted over an AWGN channel. The design SNR is used only for code construction, whereas the transmission SNR in the simulations is represented by $E_s/N_0$. 
In the simulations, $C(N,K)$ denotes the true polar code used for transmission and error evaluation. In the recognition process, only $N$ is known, while $K$ and the frozen/information-set pattern are unknown.

\subsection{Comparison with the Theoretical Bounds}

Figs.~2--7 compare the proposed DE-based bounds with the Bhattacharyya-parameter-based bounds in [22]. The results are given
for $N=64$, $N=256$, and $N=1024$, with $M=100$ and $M=500$ intercepted observations. For each setting, the code rates are
$1/4$, $1/2$, and $3/4$. The curves labeled ``Ideal'' are obtained under the SC-consistent model used in the analysis, whereas the curves
labeled ``Real'' are obtained from the practical BSC recursion. 
Here, $P_e$ denotes the recognition failure probability, i.e., the probability that the frozen/information-set pattern is not correctly recognized.

The proposed bounds are evaluated using the DE-evolved LLR densities of the synthetic channels. Specifically, the overlap
coefficient $\Omega_i$ in \eqref{eq:omega_llr} is used for the sequence-level lower bound in \eqref{eq:seq_lower}, while the
optimized Chernoff coefficient $\beta_i$ in \eqref{eq:beta_llr} is used for the upper bound in \eqref{eq:seq_upper}. 
In this way, both bounds make direct use of the LLR-density information produced by DE.
By contrast, the bounds in [22] are computed only from the Bhattacharyya parameters of the synthetic channels.

As shown in Figs.~2--7, the DE upper bound is consistently tighter than the Bhattacharyya-parameter-based upper bound. The DE lower bound
is also closer to the corresponding upper bound. For all considered values of $N$ and $M$, the gap between the DE upper and lower bounds is
within $1$ dB around $P_e=10^{-2}$.

The influence of $M$ and $N$ can also be observed from these figures. When $M$ increases from $100$ to $500$, both the bounds and the
simulation curves shift to lower SNRs, since more intercepted observations are accumulated in each position-type test. In contrast,
for a fixed $M$, increasing the code length from $N=64$ to $N=256$ and then to $N=1024$ shifts the curves to higher SNRs. This is because the
sequence-level recognition is successful only when all source-bit positions are correctly identified, 
a larger $N$ requires more source-bit positions to be recognized correctly.

The real curves are close to the ideal curves for all considered settings, which is consistent with the observation in [22]. Hence, the
results in Figs.~2--7 show that the DE-based evaluation provides a tighter characterization of the recognition error under the SC-consistent analysis model.

\begin{figure}[t]
  \centering
  \includegraphics[width=0.914\linewidth]{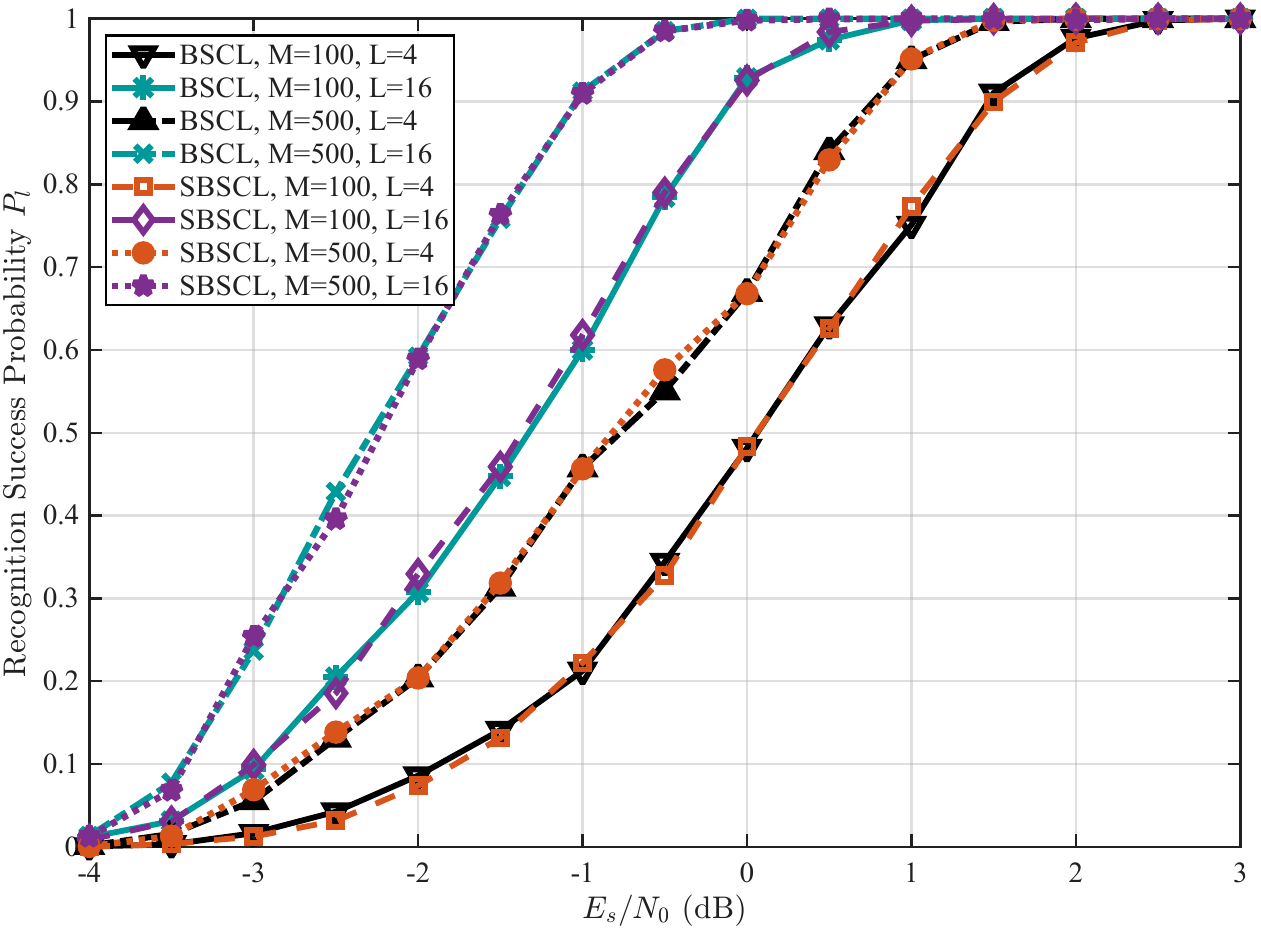}
  \caption{Performance comparison with $N=64,K=32$.}
  \label{fig:64K32}
\end{figure}

\begin{figure}[t]
  \centering
  \includegraphics[width=0.914\linewidth]{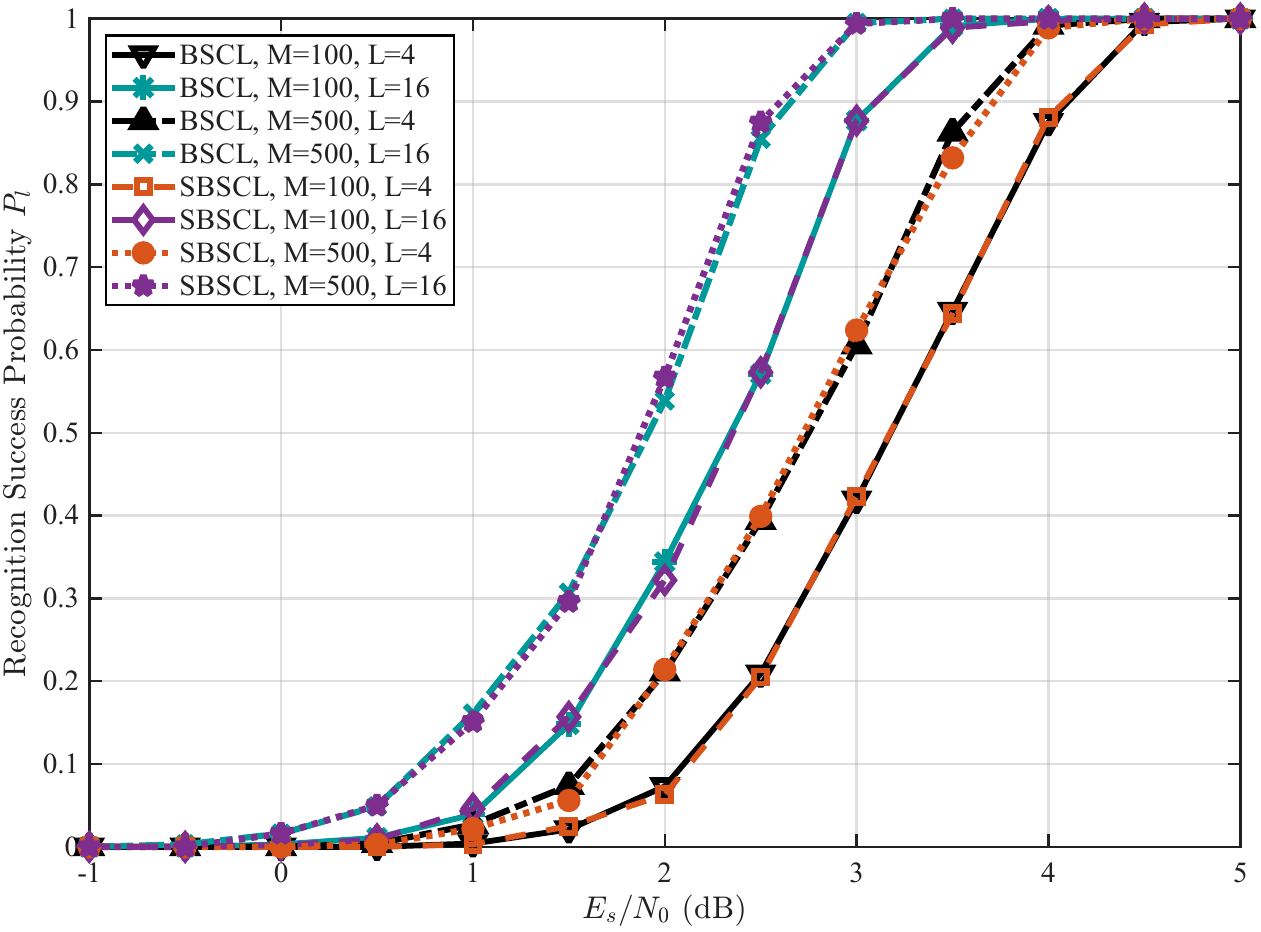}
  \caption{Performance comparison with $N=256,K=128$.}
  \label{fig:256K128}
\end{figure}

\begin{figure}[t]
  \centering
  \includegraphics[width=0.914\linewidth]{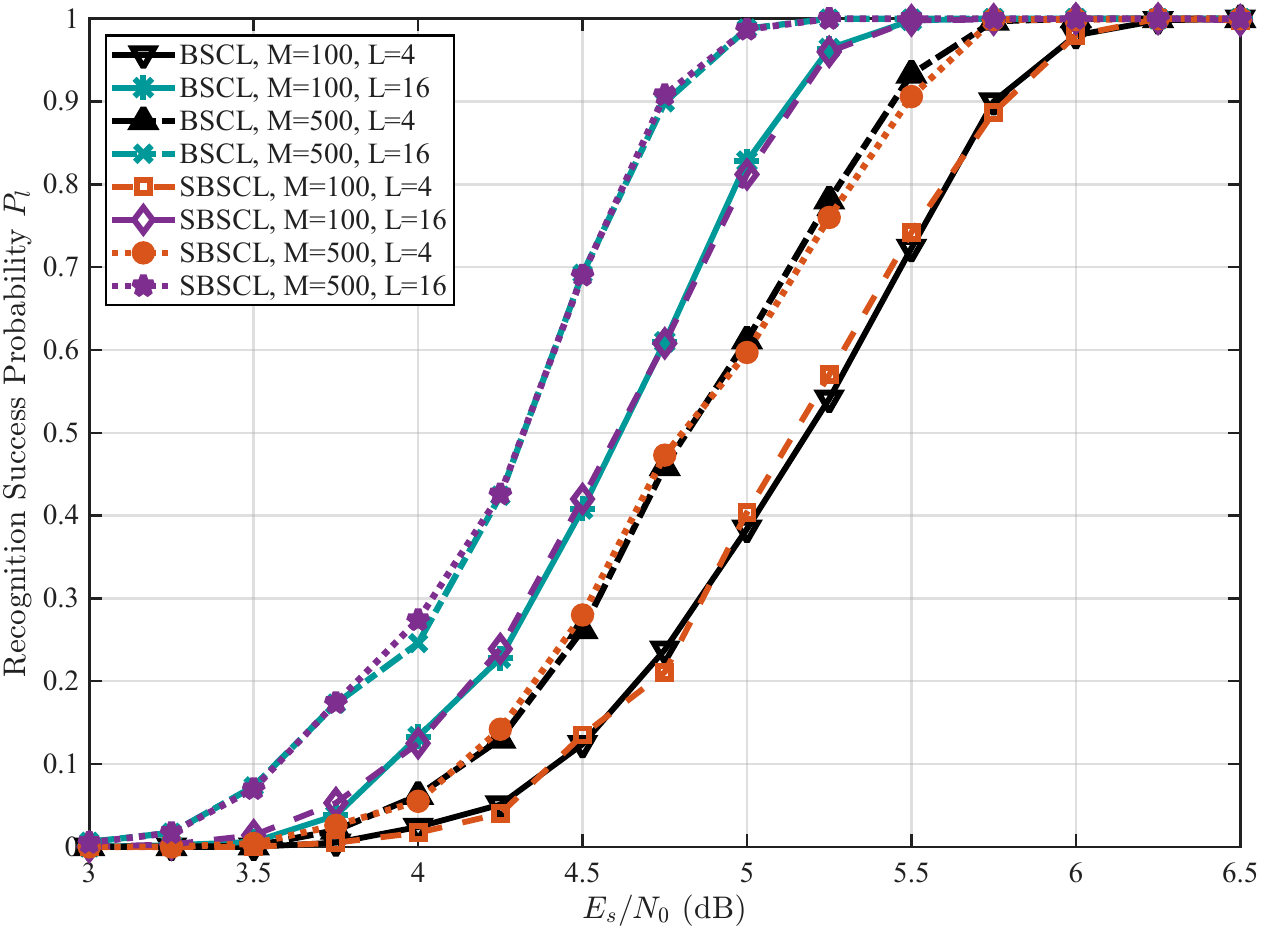}
  \caption{Performance comparison with $N=1024,K=512$.}
  \label{fig:1024K512}
\end{figure}

\begin{figure}[t]
  \centering
  \includegraphics[width=1\linewidth]{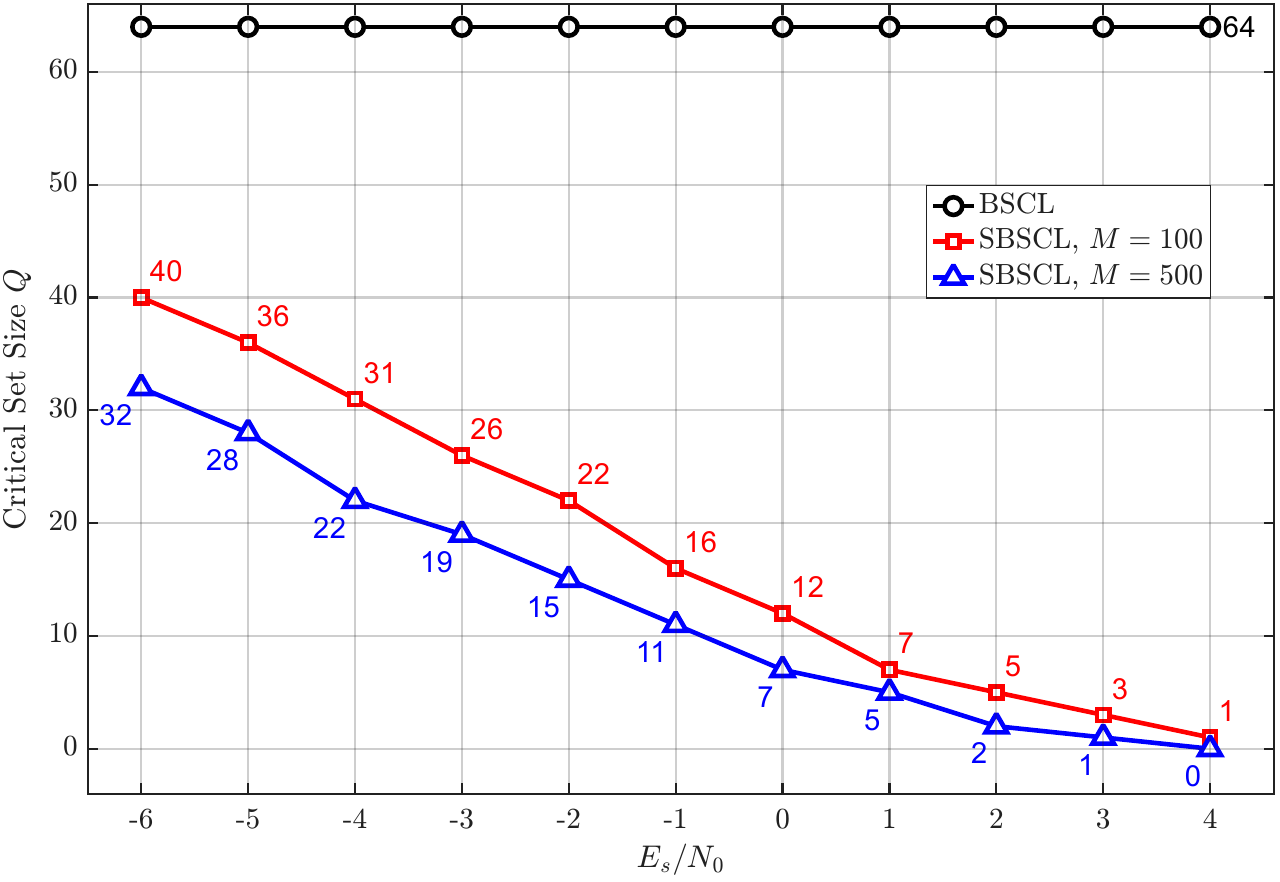}
  \caption{Critical-set size for $N=64$.}
  \label{fig:64pos}
\end{figure}

\begin{figure}[t]
  \centering
  \includegraphics[width=1\linewidth]{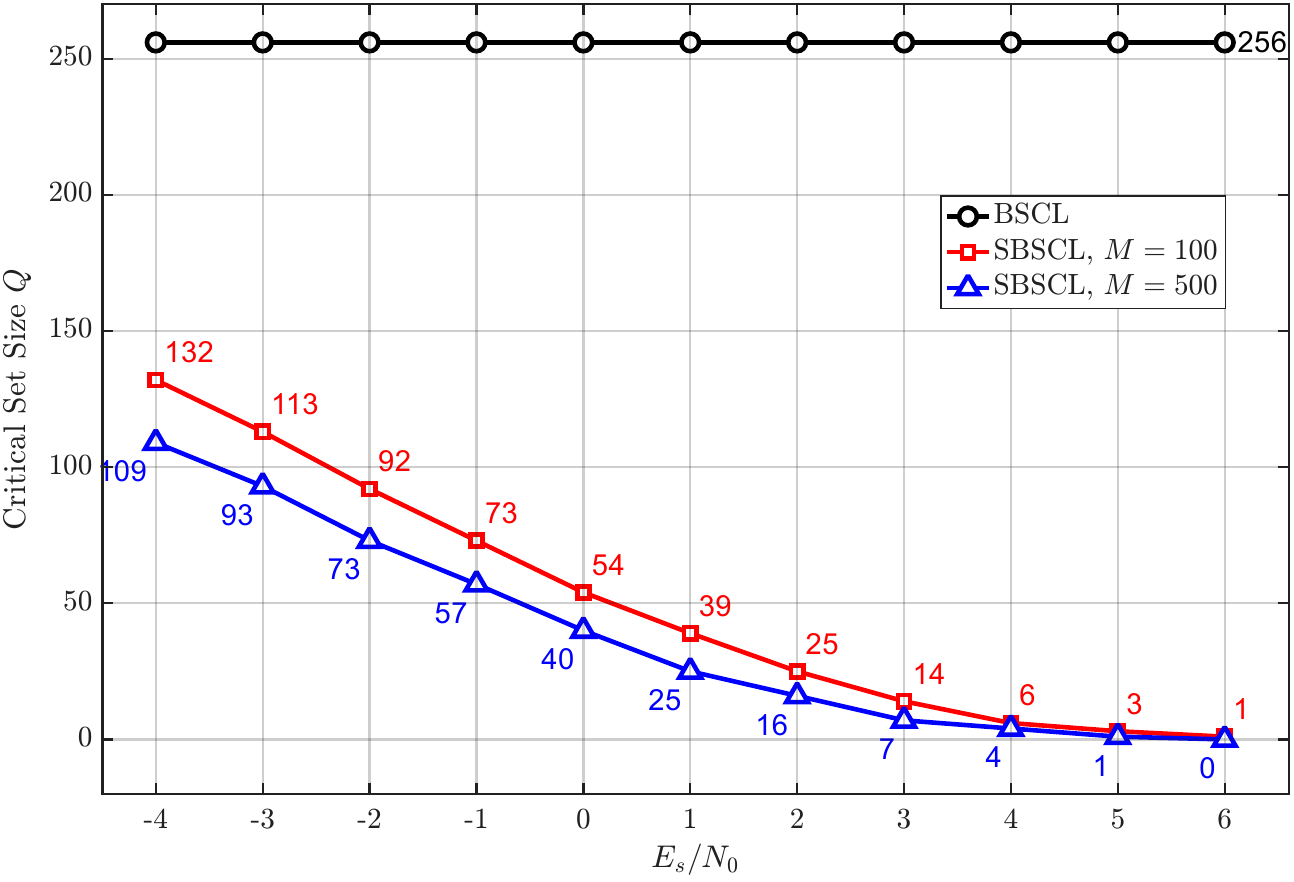}
  \caption{Critical-set size for $N=256$.}
  \label{fig:256pos}
\end{figure}

\begin{figure}[t]
  \centering
  \includegraphics[width=1\linewidth]{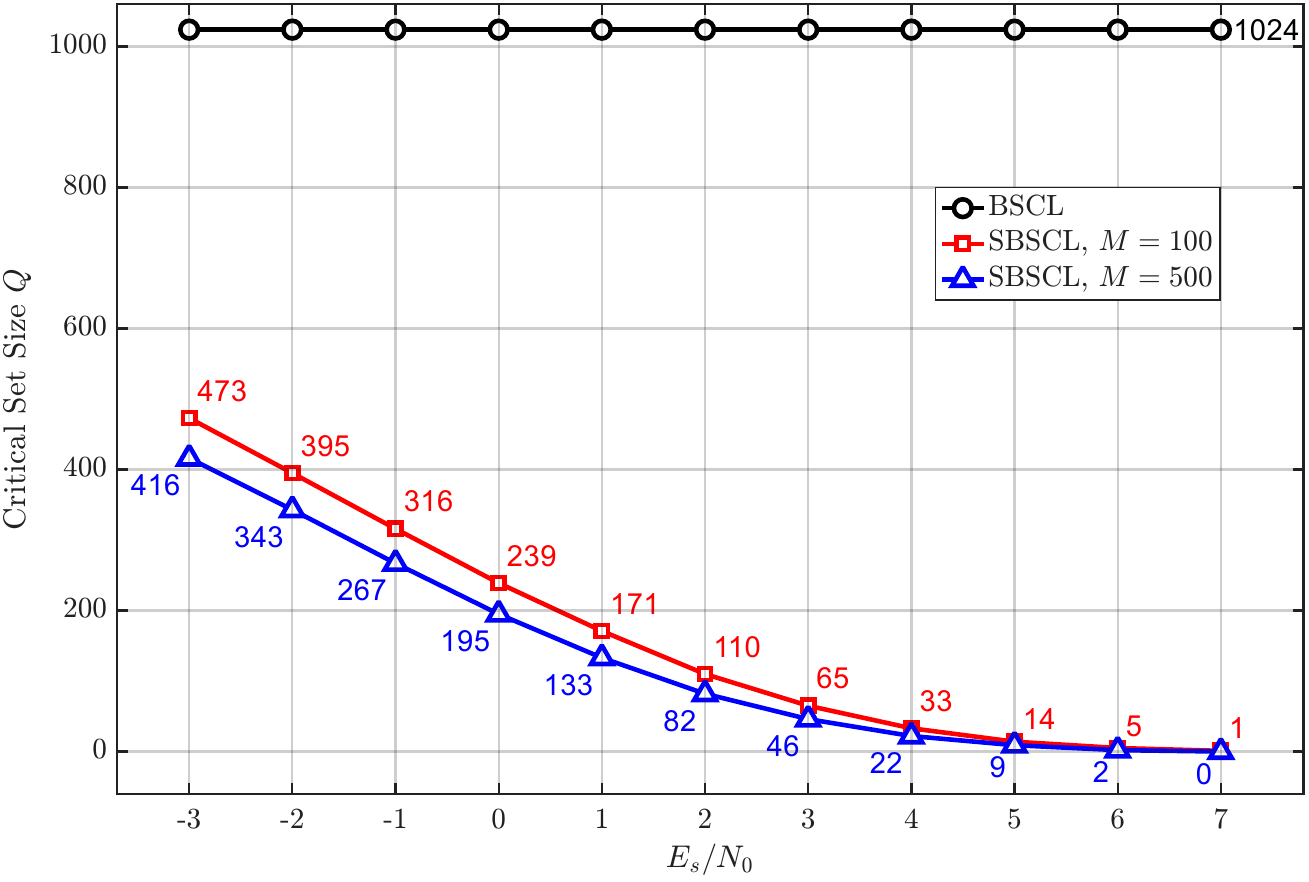}
  \caption{Critical-set size for $N=1024$.}
  \label{fig:1024pos}
\end{figure}

\subsection{Recognition Performance of BSCL and SBSCL}

Figs.~8--10 compare the recognition performance of BSCL and the proposed SBSCL. The recognition success probability is denoted by $P_l$,
which is defined as the probability that the frozen/information-set pattern is correctly recognized.

For each SNR point, $1000$ Monte Carlo trials are performed. Fig.~8 shows the results for $N=64$ and $K=32$, Fig.~9 gives the results for
$N=256$ and $K=128$, and Fig.~10 presents the results for $N=1024$ and $K=512$. In all three cases, $M=100$ and $M=500$ intercepted
observations are considered, and the list size is set to $L_{\rm list}=4$ and $L_{\rm list}=16$. 
For SBSCL, the critical set is selected according to the DE-based upper-bound contribution. The threshold is set to $\eta=10^{-2}$ in \eqref{eq:expansion_set}.

The SBSCL curves almost overlap with the BSCL curves in all simulated settings. This shows that performing the two-hypothesis list expansion
only at the positions in critical set causes little performance loss. These positions have relatively large DE-based upper-bound
contributions and are therefore more likely to affect the recognition result. The additional expansions at the remaining positions bring only limited improvement.

Several expected trends can also be observed from Figs.~8--10. Increasing $M$ improves the
recognition performance, whereas a longer code requires a higher SNR to achieve the same success probability. A larger list size also improves the
performance, since more candidate paths are retained during the recursive recognition process. These observations show that SBSCL
keeps the main list-search gain of BSCL while reducing unnecessary list operations.

\subsection{Critical-Set Size and List-Operation Reduction}

The expansion positions of SBSCL are determined by the DE-based upper-bound contribution $\beta_i^M$. The threshold is set to
$\eta=10^{-2}$ in \eqref{eq:expansion_set}. Let $Q=|\mathcal{J}_{10^{-2}}|$ denote the size of critical-set.
SBSCL performs the two-hypothesis path expansion only at these $Q$ positions, while the remaining $N-Q$ positions are processed
by a single local metric comparison. 

Figs.~11--13 show the value of $Q$ for $N=64$, $N=256$, and $N=1024$, respectively. 
It can be seen that the size of critical set decreases as the SNR increases.
This is because the position-type tests become more reliable at higher SNRs, so fewer positions satisfy the selection condition in
\eqref{eq:expansion_set}. 
Moreover, increasing $M$ can further reduce the size of critical set. This is because a larger $M$ makes the upper-bound contribution of
each position smaller, thereby improving the reliability of the local position-type test.

For $N=64$, as shown in Fig.~11, BSCL expands all $64$ positions. At $E_s/N_0=0$ dB, SBSCL selects only $12$ positions for $M=100$ and
$7$ positions for $M=500$. At higher SNRs, the size of critical set becomes very small. For example, when $M=500$, only one position is selected
at $E_s/N_0=3$ dB, and no position is selected at $E_s/N_0=4$ dB.

The same reduction can be observed for larger code lengths. For $N=256$, Fig.~12 shows that, at $E_s/N_0=0$ dB, 
the size of critical set is reduced from $256$ in BSCL to $54$ for $M=100$ and $40$ for $M=500$. When $M=500$, $Q$ further decreases to $7$, $4$, and $1$
at $E_s/N_0=3$, $4$, and $5$ dB, respectively. For $N=1024$, as shown in Fig.~13, SBSCL selects $65$ positions for $M=100$ and $46$ positions
for $M=500$ at $E_s/N_0=3$ dB, instead of expanding all $1024$ positions. At $E_s/N_0=7$ dB, $Q$ decreases to $1$ for $M=100$ and to $0$ for $M=500$.

The decrease of $Q$ directly reduces the list operations in SBSCL. In BSCL, path splitting, path-state copying, candidate sorting, and
pruning are performed at every source-bit position. In SBSCL, these operations are performed only at the positions in the critical set
$\mathcal{J}_{\eta}$. At the remaining positions, each surviving path is updated in place according to the smaller local metric increment.
This effect becomes more pronounced in the high-SNR region, where only a few positions remain in the critical set. When $Q$ is
small, the few candidate hypotheses generated at the selected positions can also be processed in parallel by several BSC recognizers. 
In the extreme case where $\mathcal{J}_{\eta}$ is empty, SBSCL reduces to BSC recognition without list operation.

\section{Conclusion}

This paper studied the simplification of BSCL recognition of polar codes. 
From the first-error statistics, we found that recognition failures are mainly concentrated at a few unreliable source-bit positions. 
This property was used to construct the critical set in SBSCL, where the two-hypothesis list expansion is kept only at the selected positions and the remaining positions are processed by a single local decision. After that, 
DE-based recognition bounds were also derived under the SC-consistent model, using the evolved synthetic LLR distributions to evaluate the frozen-bit and information-bit tests. 
The simulation results show that the DE-based bounds give a tighter evaluation than the Bhattacharyya-parameter-based bounds, and the SBSCL can reduce list operations while maintaining a recognition performance of BSCL.


\begin{thebibliography}{99}
\bibitem{1}
A.~Bonvard, S.~Houcke, R.~Gautier, and M.~Marazin,  ``Classification based on Euclidean distance distribution for blind identification of error correcting codes in noncooperative contexts,'' \emph{IEEE Trans. Signal Process.}, vol.~66, no.~10, pp.~2572--2583, May 2018.

\bibitem{2}
O.~A. Dobre, A.~Abdi, Y.~Bar-Ness, and W.~Su, ``Survey of automatic modulation classification techniques: classical approaches and new trends,'' \emph{IET Commun.}, 
vol.~1, no.~2, pp.~137--156, Apr. 2007.

\bibitem{3}
Y.-S.~Kil, H.~Lee, S.-H.~Kim, and S.-H.~Chang, ``Analysis of blind frame recognition and synchronization based on sync word periodicity,'' 
\emph{IEEE Access}, vol.~8, pp.~147516--147532, 2020.

\bibitem{4}
R.~Moosavi and E.~G. Larsson, ``Fast blind recognition of channel codes,'' \emph{IEEE Trans. Commun.}, vol.~62, no.~5, pp.~1393--1405, May 2014.

\bibitem{5}
A.~D. Yardi, S.~Vijayakumaran, and A.~Kumar, ``Blind reconstruction of binary cyclic codes from unsynchronized bitstream,'' 
\emph{IEEE Trans. Commun.}, vol.~64, no.~7, pp.~2693--2706, Jul. 2016.

\bibitem{6}
D.~Jo, S.~Kwon, and D.-J.~Shin, ``Blind reconstruction of BCH codes based on consecutive roots of generator polynomials,'' 
\emph{IEEE Commun. Lett.}, vol.~22, no.~5, pp.~894--897, May 2018.

\bibitem{7}
M.~Song, J.~Kim, and D.-J.~Shin, ``Blind reconstruction of BCH and RS codes using single-error correction,'' 
\emph{IEEE Trans. Signal Process.}, vol.~69, pp.~5120--5133, 2021.

\bibitem{8}
J.~Dingel and J.~Hagenauer, ``Parameter estimation of a convolutional encoder from noisy observations,'' 
in \emph{Proc. IEEE Int. Symp. Inf. Theory (ISIT)}, Nice, France, Jun. 2007, pp.~1776--1780.

\bibitem{9}
R.~Swaminathan and A.~S. Madhukumar, ``Blind parameter estimation of turbo convolutional codes: Noisy and non-synchronized scenario,''
\emph{Digit. Signal Process.}, vol.~95, Art. no.~102577, Dec. 2019.

\bibitem{10}
T.~Xia and H.-C.~Wu, ``Novel blind identification of LDPC codes using average LLR of syndrome a posteriori probability,'' 
\emph{IEEE Trans. Signal Process.}, vol.~62, no.~3, pp.~632--640, Feb. 2014.

\bibitem{11}
Z.~Wu, L.~Zhang, Z.~Zhong, and R.~Liu, ``Blind recognition of LDPC codes over candidate set,'' 
\emph{IEEE Commun. Lett.}, vol.~24, no.~1, pp.~11--14, Jan. 2020.

\bibitem{12}
E. Ar{\i}kan, “Channel polarization: A method for constructing capacity-achieving codes for symmetric binary-input memoryless channels,” \emph{IEEE Trans. Inf. Theory}, vol. 55, no. 7, pp. 3051--3073, Jul. 2009.

\bibitem{13}
K. Niu and K. Chen, “CRC-aided decoding of polar codes,” \emph{IEEE Commun. Lett.}, vol. 16, no. 10, pp. 1668–1671, Oct. 2012.

\bibitem{14}
3rd Generation Partnership Project (3GPP), ``NR; Multiplexing and channel coding,'' TS 38.212, Rel. 15, Jul. 2018.

\bibitem{15}
Q.~Liu, H.~Zhang, and P.~Yu,
``Blind detection and identification of polar codes based on the parity check conformity,''
\emph{IEEE Commun. Lett.},
vol.~26, no.~4, pp.~728--732, Apr. 2022.

\bibitem{16}
J. Liu, T. Zhang, H. Bai, and S. Ye, “Blind recognition algorithm of polar code based on information matrix estimation,” \emph{Syst. Eng. Electron.}, vol. 44, no. 2, pp. 668--676, 2022.

\bibitem{17}
C. Yi, B. Pang, L. He, B. Ma, Y. Li, and F. C. M. Lau, “Blind identification of polar codes based on estimation and derivation approaches,” \emph{IEEE Commun. Lett.}, vol. 27, no. 2, pp. 414--418, Feb. 2023.

\bibitem{18}
P. Xu, J. Liu, A. Wang, C. Yi, and Q. Li, “Blind recognition of polar code information bits based on multi-threshold voting and partial orders,” \emph{IEEE Commun. Lett.}, vol. 30, pp. 887--891, 2026.

\bibitem{19}
Z. Wu, Z. Zhong, L. Zhang, and B. Dan, “Recognition of non-drilled polar codes based on soft decision,” \emph{J. Commun.}, vol. 41, no. 12, pp. 60--71, Dec. 2020.

\bibitem{20}
Y. Wang, C. Wang, X. Wang, and Z. Huang, “Non-punctured polar code parameter recognition algorithm based on soft decision,” \emph{Syst. Eng. Electron.}, vol. 45, no. 10, pp. 3293--3301, Oct. 2023.

\bibitem{21}
C. Tu, Y. Liu, X. Feng, and K. Niu, ``Blind recognition of polar codes using successive cancellation list decoding,'' \emph{arXiv preprint arXiv:2605.13331}, May 2026.

\bibitem{22}
C.~Tu, C.~Yang, X.~Feng, and K.~Niu, ``A hypothesis-testing analysis of blind recognition for polar codes,'' \emph{arXiv preprint arXiv:2606.17705}, Jun. 2026.

\bibitem{23}
H. L. Van Trees, \emph{Detection, Estimation, and Modulation Theory, Part I}. New York, NY, USA: Wiley, 2001.


\bibitem{24}	
P. Trifonov, “Efficient design and decoding of polar codes,” \emph{IEEE Trans. Commun.}, vol. 60, no. 11, pp. 3221–3227, Nov. 2012.

\end{thebibliography}
\end{document}